\documentclass[10pt,aps,prl,twocolumn,notitlepage,showpacs,superscriptaddress]{revtex4-1}
\usepackage{amsthm,amsmath,amsfonts,amssymb,verbatim,color}
\usepackage{graphicx,times}
\usepackage{bm}
\usepackage{epsfig,slashed}
\usepackage{amssymb}
\usepackage{amsfonts}
\usepackage{tikz}
\usepackage{xparse}
\usepackage{ifthen}
\usepackage{lipsum}
\usepackage{rotating}
\usepackage[colorlinks=true,citecolor=blue,
linkcolor=blue,urlcolor=blue]{hyperref}
\usepackage[caption=false]{subfig}
\begin{document}

\newcommand{\tr}{\mathop{\mathrm{tr}}}
\newcommand{\bsigma}{\boldsymbol{\sigma}}
\newcommand{\re}{\mathop{\mathrm{Re}}}
\newcommand{\im}{\mathop{\mathrm{Im}}}
\renewcommand{\b}[1]{{\boldsymbol{#1}}}
\renewcommand{\c}[1]{\mathcal{#1}}
\newcommand{\diag}{\mathrm{diag}}
\newcommand{\sign}{\mathrm{sign}}
\newcommand{\sgn}{\mathop{\mathrm{sgn}}}

\newcommand{\cl}{\mathrm{cl}}
\newcommand{\mb}{\bm}
\newcommand{\ua}{\uparrow}
\newcommand{\da}{\downarrow}
\newcommand{\ra}{\rightarrow}
\newcommand{\la}{\leftarrow}
\newcommand{\mc}{\mathcal}
\newcommand{\bs}{\boldsymbol}
\newcommand{\lra}{\leftrightarrow}
\newcommand{\nn}{\nonumber}
\newcommand{\half}{{\textstyle{\frac{1}{2}}}}
\newcommand{\mf}{\mathfrak}
\newcommand{\MF}{\text{MF}}
\newcommand{\IR}{\text{IR}}
\newcommand{\UV}{\text{UV}}
\newcommand{\sech}{\mathrm{sech}}

\title{Triangular pair density wave in confined superfluid $^3$He}

\author{Pramodh Senarath Yapa}
\affiliation{Department of Physics, University of Alberta, Edmonton, Alberta T6G 2E1, Canada}

\author{Rufus Boyack}
\affiliation{Department of Physics, University of Alberta, Edmonton, Alberta T6G 2E1, Canada}
\affiliation{Theoretical Physics Institute, University of Alberta, Edmonton, Alberta T6G 2E1, Canada}

\author{Joseph Maciejko}
\affiliation{Department of Physics, University of Alberta, Edmonton, Alberta T6G 2E1, Canada}
\affiliation{Theoretical Physics Institute, University of Alberta, Edmonton, Alberta T6G 2E1, Canada}

\date\today

\begin{abstract}
Recent advances in experiment and theory suggest that superfluid $^3$He under planar confinement may form a pair density wave (PDW) whereby superfluid and crystalline orders coexist. While a natural candidate for this phase is a unidirectional stripe phase predicted by Vorontsov and Sauls in 2007, recent nuclear magnetic resonance measurements of the superfluid order parameter rather suggest a two-dimensional PDW with noncollinear wavevectors, of possibly square or hexagonal symmetry. In this Letter, we present a general mechanism by which a PDW with the symmetry of a triangular lattice can be stabilized, based on a superfluid generalization of Landau's theory of the liquid-solid transition. A soft-mode instability at finite wavevector within the translationally invariant planar-distorted B phase triggers a transition from uniform superfluid to PDW that is first order due to a cubic term generally present in the PDW free-energy functional. This cubic term also lifts the degeneracy of possible PDW states in favor of those for which wavevectors add to zero in triangles, which in two dimensions uniquely selects the triangular lattice.
\end{abstract}

\maketitle

{\it Introduction.---}$^3$He arguably best epitomizes the paradigm of emergence in condensed matter physics. Atomically one of the simplest isotopes in the periodic table of elements, $^3$He nonetheless gives rise to a rich variety of paired superfluid phases at low temperatures. In bulk and absent magnetic fields, only two superfluid phases are thermodynamically stable~\cite{VW3He}: the B phase with isotropic quasiparticle gap appears at zero pressure, and an additional A phase with nodal quasiparticles is stabilized at high pressure. In contrast to $^4$He atoms or $s$-wave Cooper pairs, the $p$-wave Cooper pairs in $^3$He couple strongly to geometric perturbations on account of their spatial anisotropy, leading to the possibility of new superfluid phases under confinement.

While superfluid $^3$He in planar confinement was first studied decades ago~\cite{freeman1988,freeman1990}, recent developments have led to a resurgence of interest in the subject. For $^3$He confined to a slab of thickness $D$ on the order of the superfluid coherence length $\xi_0$, Ginzburg-Landau (GL)~\cite{li1988} and quasiclassical~\cite{hara1988,nagato1998,nagato2000,vorontsov2003} theories predict that the A phase may appear at zero pressure, and that the B phase gives way to the planar-distorted B phase (PDB phase) with a gap that differs in directions parallel and normal to the confinement plane. Owing to advances in microfabrication techniques, those early predictions have recently been verified experimentally~\cite{levitin2013,levitin2013b,zhelev2017,levitin2019,shook2020}. Since for moderate confinement the PDB phase can be viewed as a three-dimensional (3D) topological superfluid~\cite{schnyder2008,roy2008,qi2009,volovik2009b}, such $^3$He films would also provide an ideal platform for the detection of Majorana fermions~\cite{volovik2009,chung2009,wu2013,park2015}.

In addition to the A and PDB phases, which are homogeneous in the confinement plane, Vorontsov and Sauls predicted in 2007 that confinement could lead to an additional superfluid phase with spontaneously broken translation symmetry---the \emph{stripe phase}~\cite{vorontsov2007,wiman2016,aoyama2016}. The stripe phase is a unidirectional pair density wave (PDW)~\cite{agterberg2020} analogous to the Larkin-Ovchinnikov state~\cite{larkin1965,radzihovsky2009,radzihovsky2011} that can also be understood as a periodic arrangement of domain walls~\cite{salomaa1988,silveri2014} in the PDB phase order parameter. While such domain walls are not energetically favorable in bulk $^3$He, they reduce the amount of surface pair-breaking relative to a homogeneous phase under planar confinement~\cite{vorontsov2005}.

Recent experiments have sought evidence of the stripe phase via nuclear magnetic resonance (NMR) measurements of the superfluid order parameter~\cite{levitin2019} and fourth-sound measurements of the superfluid density~\cite{shook2020}. The latter experiment suggested that a new phase sandwiched between the A and PDB phases appears under sufficient confinement, although the precise nature of this phase and of the transitions surrounding it remains to be elucidated. Reference~\cite{levitin2019} likewise found evidence of a new phase in the vicinity of the A-PDB transition, but the NMR signatures of this phase were inconsistent with the stripe phase. Specifically, the observation of a kink in the NMR frequency shift at a critical tipping angle $\beta^*$ ruled out the stripe phase, which should exhibit no such kink~\cite{wiman2016}, but the measured value of $\beta^*$ did not match that expected for the translation-invariant PDB phase. Rather, the authors of Ref.~\cite{levitin2019} reconciled those observations by proposing a ``polka-dot phase''~\cite{saunders2018}: a \emph{two-dimensional} (2D) PDW, of possibly hexagonal or square symmetry. The possibility of a 2D PDW as an alternative to the 1D stripe phase was recognized by Vorontsov and Sauls~\cite{vorontsov2007,vorontsov2018}, but not explored quantitatively~\cite{Note2d}. A preliminary GL analysis for a 2D PDW with the symmetry of a square lattice, mentioned in Ref.~\cite{levitin2019}, found that such a phase was only metastable, with a free energy higher than that of the stripe phase.

To resolve this conundrum from a theoretical standpoint, we provide in this Letter a general physical argument---summarized in this paragraph and later substantiated by explicit calculations---according to which a PDW in confined $^3$He ought indeed to be two-dimensional, and also to possess the hexagonal symmetry of a {\it triangular} lattice. Our theory can be viewed as a generalization of Landau's theory of weak crystallization~\cite{landau1937} to superfluids. In Landau's theory, the Fourier component $n_\b{q}$ of an equilibrium deviation $\delta n(\b{r})\!=\!n(\b{r})\!-\!n_0$ of a classical fluid's density $n(\b{r})$ from uniform density $n_0$ serves as the order parameter for the liquid-solid transition. As the transition is approached from the liquid side, density fluctuations become strongly peaked in reciprocal space on a surface of momenta $|\b{q}|\!=\!Q$ which is spherical on account of the fluid's unbroken rotational symmetry. Although this naively induces an instability towards crystallization for a continuously degenerate set of wavevectors~\cite{brazovskii1975}, the Landau free-energy functional contains a cubic term $\sim\sum_{\b{q}_1,\b{q}_2,\b{q}_3}n_{\b{q}_1}n_{\b{q}_2}n_{\b{q}_3}\delta_{\b{q}_1+\b{q}_2+\b{q}_3,0}$ which lifts this degeneracy, simultaneously rendering the transition first order and favoring crystalline lattices for which wavevectors add up to zero in triangles. A first attempt to transpose these ideas to a PDW or ``superfluid crystallization'' transition in $^3$He meets the objection that cubic terms in the GL functional for a superfluid are forbidden by $U(1)$ gauge symmetry. The latter functional is however appropriate for a superfluid transition out of the normal state. As we show below, the Landau functional for a PDW transition {\it within} a superfluid state, in which gauge symmetry is already broken, can indeed contain a cubic term. This term is analogous to that of the liquid-solid transition, with $n_\b{q}$ replaced by the Fourier components of the superfluid order parameter. Combined with the softening at a finite wavevector $|\b{q}|\!=\!Q$ of a particular collective mode in the confined superfluid~\cite{mizushima2018}, the cubic term drives a first-order transition from the homogeneous superfluid to a PDW whose Bravais lattice structure, being two-dimensional, is necessarily triangular.

{\it Fluctuations in the PDB phase.---}Our starting point is the GL free-energy functional for $^3$He~\cite{VW3He},
\begin{widetext}
\begin{align}\label{F}
    F=\int d^3r\Bigl[&K_1\partial_k A_{\mu j}\partial_k A_{\mu j}^*
+K_2\partial_j A_{\mu j}\partial_k A_{\mu k}^*
+K_3\partial_k A_{\mu j}\partial_j A_{\mu k}^*
+\alpha \tr AA^\dag\nn\\
&+\beta_1|\tr AA^T|^2+\beta_2(\tr AA^\dag)^2
+\beta_3 \tr AA^T(AA^T)^*+\beta_4\tr(AA^\dag)^2
+\beta_5\tr AA^\dag(AA^\dag)^*\Bigr],
\end{align}
\end{widetext}
where $A_{\mu j}$ is the $3\!\times\!3$ superfluid order parameter with $\mu$ spin and $j$ orbital indices, and $\alpha$, $\beta_1,\ldots,\beta_5$, and $K_1,K_2,K_3$ are phenomenological parameters. We will employ values of these parameters given by the weak-coupling approximation~\cite{VW3He}, but comment on the effect of strong-coupling corrections~\cite{choi2007,wiman2016} at the end. Assuming planar confinement with specular interfaces in the $z$ direction, the order parameter in the PDB phase is~\cite{vorontsov2003}:
\begin{align}\label{PDB}
    \overline{A}_{\mu j}(z)=\left(\begin{array}{ccc}
\Delta_\parallel(z) & 0 & 0 \\
0 & \Delta_\parallel(z) & 0 \\
0 & 0 & \Delta_\perp(z)
\end{array}\right),
\end{align}
which incorporates the planar phase~\cite{vorontsov2003} with $\Delta_\perp(z)\!=\!0$ as a special case. The order parameter is uniform in the confinement ($xy$) plane and invariant under simultaneous $SO(2)_{L_z+S_z}$ rotations of the orbital and spin coordinates about the $z$ axis. To search for a PDW instability in the PDB phase, we write $A_{\mu j}$ as:
\begin{align}\label{expansion}
    A_{\mu j}(\b{r}_\parallel,z)=\overline{A}_{\mu j}(z)+\sum_\b{q}\phi_{\mu j}(\b{q},z)e^{i\b{q}\cdot\b{r}_\parallel},
\end{align}
where $\b{r}_\parallel\!=\!(x,y)$ and $\b{q}\!=\!(q_x,q_y)$. Our strategy is to construct a Landau functional for the fluctuation $\phi$ by treating it as a small correction to the PDB order parameter (\ref{PDB}) and expanding (\ref{F}) to quartic order in $\phi$. Subtracting the free energy $F[\overline{A}]$ of the PDB phase, and dividing by the sample area, the resulting PDW free-energy-density functional is of the form:
\begin{align}\label{FPDW}
    f_\text{PDW}[\phi]=f^{(2)}[\phi]+f^{(3)}[\phi]+f^{(4)}[\phi],
\end{align}
where the superscripts refer to the order of expansion in $\phi$, and linear terms are absent since $\overline{A}$ is a stationary point of $F$.

We first focus on the quadratic term $f^{(2)}$, which contains contributions from both the quadratic and quartic terms in (\ref{F}), and can be written as:
\begin{align}\label{f2}
    f^{(2)}[\phi]=\sum_\b{q}\int_0^Ddz\,\phi^*_{\mu j}(\b{q},z)\hat{C}_{\mu j,\nu k}(\b{q},z)\phi_{\nu k}(\b{q},z),
\end{align}
where $D$ is the sample thickness, and $\hat{C}$ is a $\b{q}$-dependent Hermitian differential operator that can be written as the sum of two terms, $\hat{C}(\b{0},z)\!+\!\delta\hat{C}(\b{q},z)$. $\hat{C}(\b{0},z)$ contains ``kinetic'' terms proportional to $\partial_z^2$ as well as $z$-dependent ``potential'' terms quadratic in the equilibrium order parameters $\Delta_\parallel(z)$ and $\Delta_\perp(z)$~\cite{mizushima2018}. $\delta\hat{C}(\b{q},z)$ contains explicitly $\b{q}$-dependent terms arising from the derivative terms in (\ref{F}).

\begin{figure}[t]
    \includegraphics[width=\columnwidth]{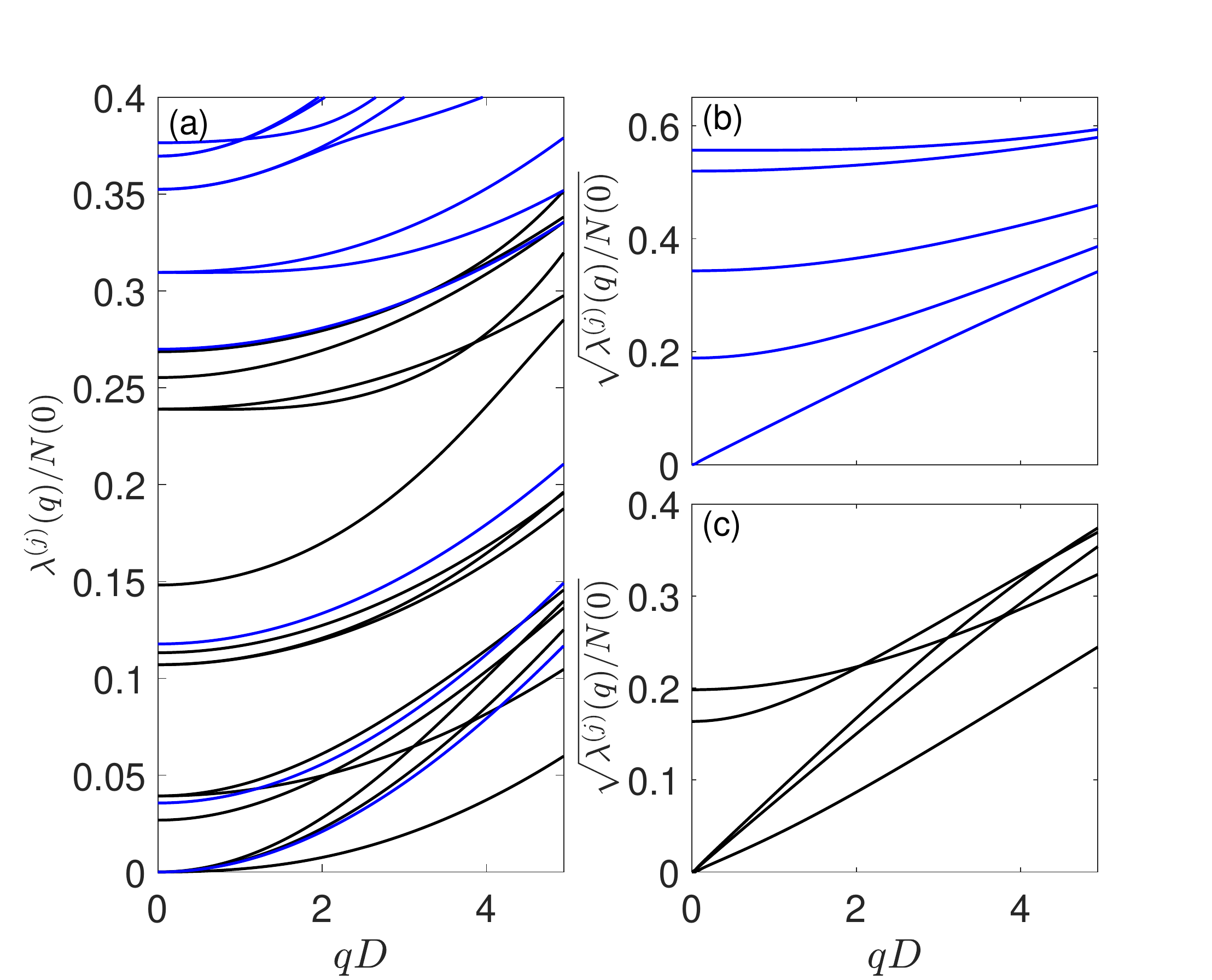}
    \caption{(a) Eigenvalues $\lambda^{(j)}(\b{q})$ in units of the normal-state density of states $N(0)$ vs momentum $q\!=\!|\b{q}|$ for real (black) and imaginary (blue) normal modes in the PDB phase; their square root (b,c) is proportional to collective mode frequencies. Parameters are chosen as $D\!=\!300$~nm, $P\!=\!10$~bar, and $T\!=\!0.914$~mK; other parameters in the PDB phase give qualitatively similar results.}
    \label{fig:Eigs}
    \centering
\end{figure}

In the spirit of a Landau expansion, we first use Eq.~(\ref{f2}) to determine the normal modes of the system and isolate the particular mode that becomes critical at the PDW transition. These normal modes are the eigenvectors $\Phi_\b{q}^{(j)}(z)$ of $\hat{C}(\b{q},z)$ with eigenvalues $\lambda^{(j)}(\b{q})$, which we compute numerically~\cite{suppmat}. Expanding the fluctuation $\phi$ in (\ref{f2}) in terms of those normal modes, we have:
\begin{align}\label{f2lambda}
    f^{(2)}[\phi]=\sum_\b{q}\sum_j\lambda^{(j)}(\b{q})|u^{(j)}_\b{q}|^2,
\end{align}
where $u^{(j)}_\b{q}$ is the amplitude of the fluctuation in the normal mode $\Phi_\b{q}^{(j)}(z)$. In Fig.~\ref{fig:Eigs}(a), we plot the lowest 25 eigenvalues, which by rotational invariance in the PDB phase depend only on the magnitude $q$ of the wavevector $\b{q}$. Omitting spin and orbital indices, the fluctuation $\phi(\b{r})\!=\!\sum_\b{q}\phi(\b{q},z)e^{i\b{q}\cdot\b{r}_\parallel}$ can be decomposed into real $\phi^+(\b{r})$ and imaginary $\phi^-(\b{r})$ parts which do not mix at quadratic order because of time-reversal symmetry in the PDB phase; the normal modes can then be separated into real and imaginary modes according to this decomposition. At $\b{q}\!=\!0$, the normal modes carry an additional $SO(2)_{L_z+S_z}$ angular momentum quantum number $m\!=\!0,\pm 1,\pm 2$~\cite{mizushima2018}; a nonzero $\b{q}$ acts as a vector perturbation which mixes $\b{q}\!=\!0$ eigenmodes with different angular momenta. By plotting the square root $\sqrt{\lambda^{(j)}(\b{q})}$ of the normal mode eigenvalues, which is proportional to bosonic collective mode frequencies $\omega_j(\b{q})$~\cite{mizushima2018}, we find one imaginary [Fig.~\ref{fig:Eigs}(b)] and three real [Fig.~\ref{fig:Eigs}(c)] linearly dispersing Goldstone modes. The former corresponds to the usual $U(1)$ phase (phonon) mode of neutral superfluids, and the latter are associated with the $SO(3)_S\times SO(2)_{L_z}\!\rightarrow\!SO(2)_{L_z+S_z}$ breaking of spin and orbital symmetries peculiar to the PDB phase~\cite{suppmat}.

\begin{figure}[t]
    \includegraphics[width=\columnwidth]{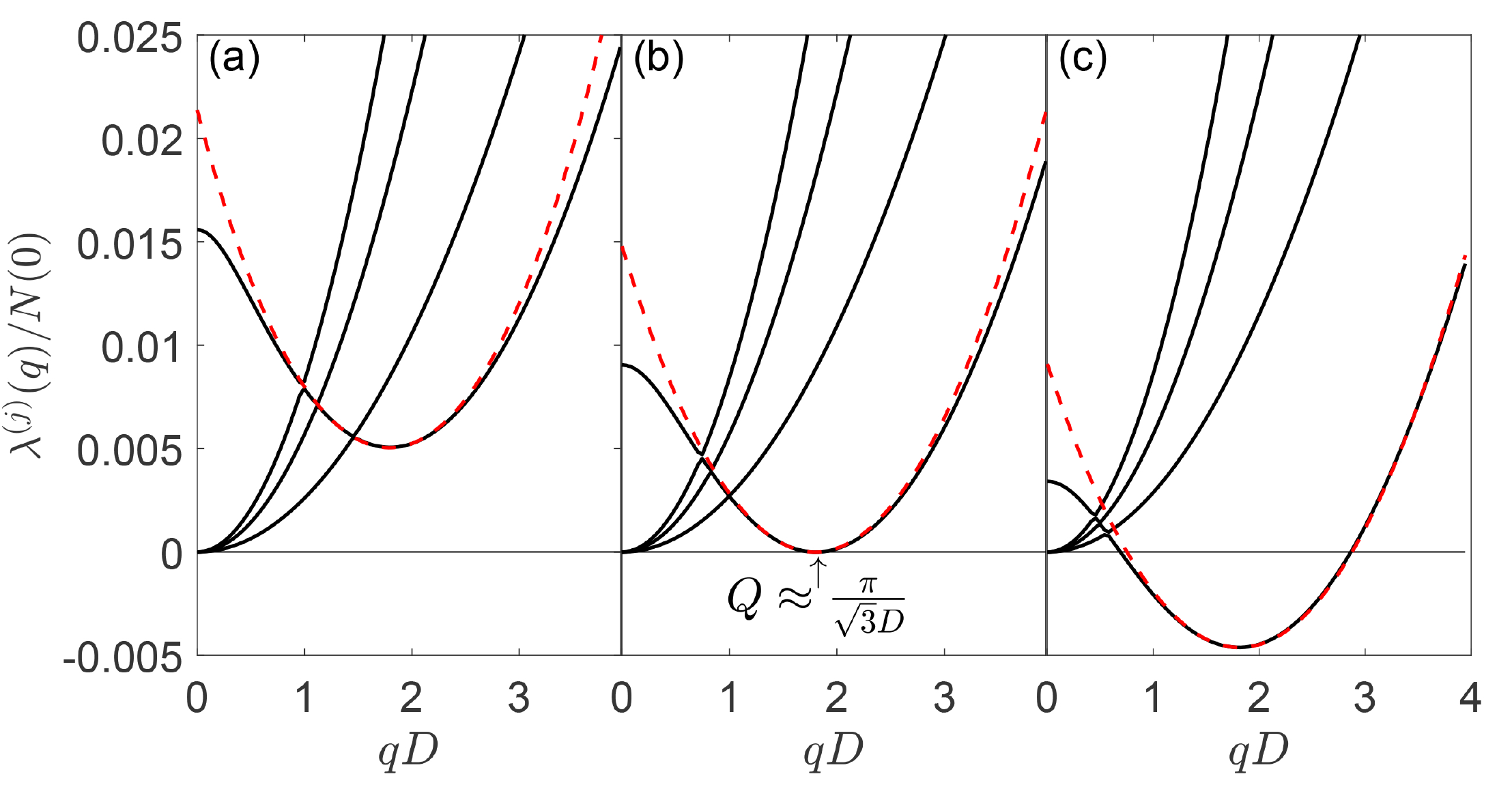}
    \caption{Black lines: eigenvalues $\lambda^{(j)}(\b{q})$ vs momentum $q\!=\!|\b{q}|$ for real normal modes in the vicinity of the softening instability at $T\!=\!T_1^*$: (a) before softening ($T\!=\!1.318$~mK, $r>0$); (b) at softening ($T\!=\!T_1^*\!=1.336$~mK, $r\!=\!0$); (c) after softening ($T\!=\!1.380$~mK, $r\!<\!0$). Here $r$ is the tuning parameter for the instability, appearing in the approximate form $\lambda^{(j_*)}(\b{q})\approx r+\kappa(\b{q}^2-Q^2)^2$ of the critical mode eigenvalue (dashed red line). Parameters are chosen as $D\!=\!300$~nm and $P\!=\!10$~bar, and we find $Q\approx\pi/(\sqrt{3}D)$ as in Ref.~\cite{wiman2016}.}
    \label{fig:Softening}
    \centering
\end{figure}

\begin{figure}[t]
    \includegraphics[width=\columnwidth]{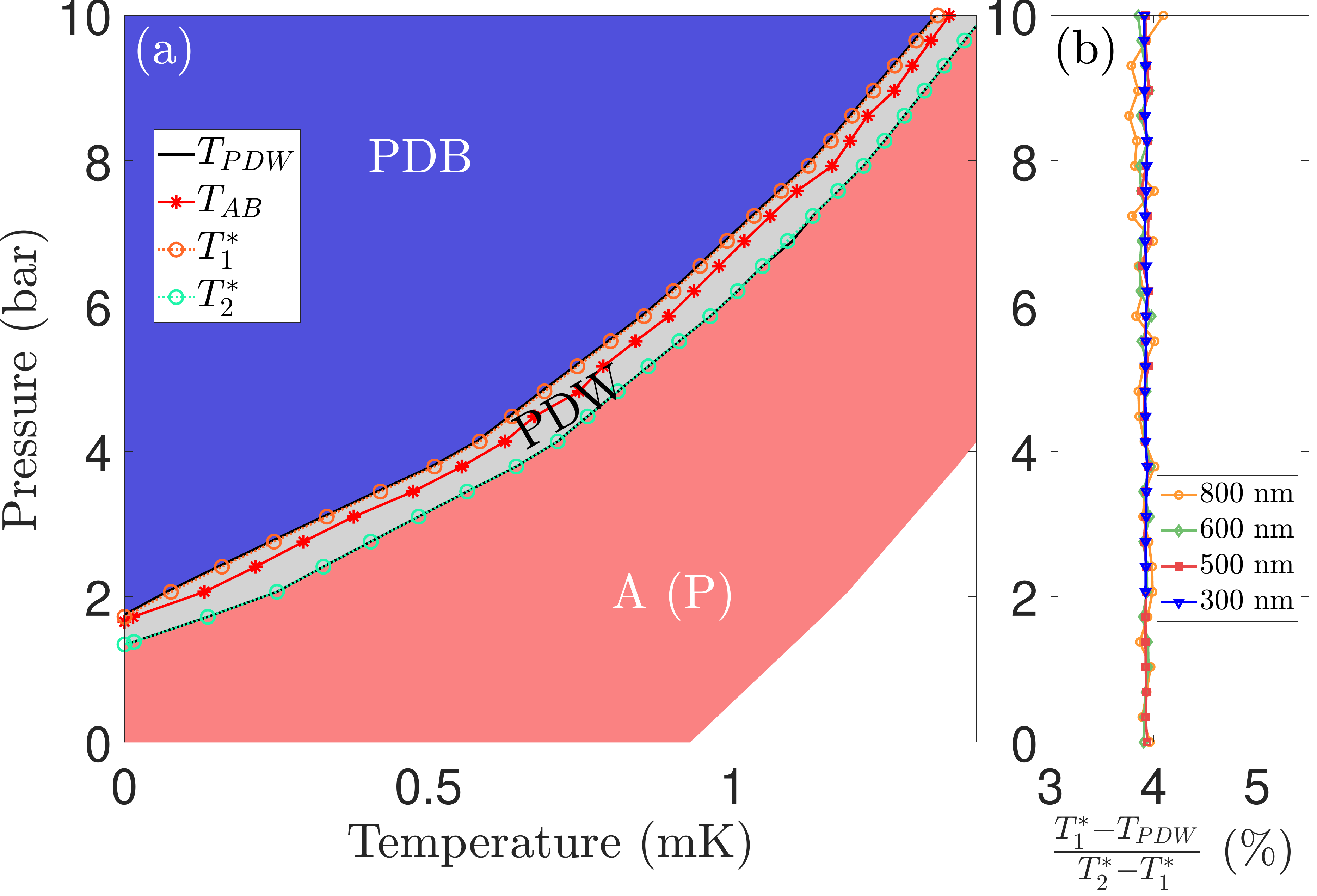}
        \caption{(a) Phase diagram of a $D\!=\!300$~nm slab, with mode-softening temperatures $T_1^*$ and $T_2^*$ flanking the would-be A-PDB transition line $T_\text{AB}$. The instability on the PDB side ($T_1^*$) is pre-empted by a first-order transition at $T_\text{PDW}$, which (b) is approximately 4\% lower than $T_1^*$ relative to the width of the instability region.}
    \label{fig:phasediagram}
    \centering
\end{figure}

{\it Mode softening.---}In addition to the gapless Goldstone modes, the PDB phase supports gapped collective modes which, should they soften as external parameters such as temperature $T$, pressure $P$, or confinement $D$ are varied, can lead to additional symmetry-breaking instabilities within the superfluid state. Such mode softening was observed in Ref.~\cite{mizushima2018}, where upon tuning the sample thickness across a certain critical value, the frequency $\omega_{j_*}(\b{q})$ of a particular real collective mode $j\!=\!j_*$ was found to touch zero at a finite wavevector $|\b{q}|\!=\!Q$ and subsequently become purely imaginary. The ensuing instability was then argued to lead to a 1D stripe phase with wavevector $Q$. In our GL description (\ref{f2lambda}), this softening corresponds to the (real) eigenvalue $\lambda^{(j_*)}(\b{q})$ for a particular real normal mode $\Phi_\b{q}^{(j_*)}(z)$ continuously crossing from positive to negative on a ring of momenta $|\b{q}|\!=\!Q$ (Fig.~\ref{fig:Softening}), and identifies this mode as the critical mode for the PDW transition. The instability region is bounded by two mode softening temperatures $T_1^*$ and $T_2^*$ that straddle the A-PDB transition line [Fig.~\ref{fig:phasediagram}(a)]; given our expansion (\ref{expansion}), $T_2^*$ corresponds here to an instability of the planar (P) phase, which is degenerate with the A phase at weak coupling~\cite{vorontsov2003}. Discarding the noncritical modes in the vicinity of the transition, we can thus approximate:
\begin{align}\label{f2PDW}
    f^{(2)}[\phi]\approx\sum_\b{q}\left[r+\kappa(\b{q}^2-Q^2)^2\right]|u_\b{q}|^2,
\end{align}
where $u_\b{q}\equiv u^{(j_*)}_\b{q}$ is the critical mode amplitude, which obeys $u_\b{q}^*=u_{-\b{q}}$ since $\phi^+(\b{r})$ and its normal-mode decomposition are real. We have also replaced the exact normal mode eigenvalue $\lambda^{(j_*)}(\b{q})$ by an approximate functional form which captures its key qualitative features near the instability (Fig.~\ref{fig:Softening}). The parameter $r$ changes sign across the instability, and $\kappa$ controls the velocity of the linearly dispersing mode $\omega_{j_*}(\b{q})\!\propto\!\sqrt{\kappa}Q\bigl||\b{q}|\!-\!Q\bigr|$ that obtains at criticality ($r\!=\!0$) near $|\b{q}|\!=\!Q$. As we now discuss, this further approximation---while not strictly necessary---allows for a simplified analytical treatment of the PDW transition that makes its analogy to the liquid-solid transition manifest~\cite{ChaikinLubensky}.

{\it Landau theory of the PDW transition.---}Our analysis so far has identified the critical normal-mode amplitude $u_\b{q}$---or equivalently its position-space inverse Fourier transform $u(\b{r}_\parallel)\!=\!\sum_\b{q}e^{i\b{q}\cdot\b{r}_\parallel}u_\b{q}$, which is a real function---as the appropriate order parameter for the PDW transition. A well-behaved Landau functional for $u$ must include terms beyond quadratic order, i.e., self-interaction terms which arise from the terms cubic and quartic in $\phi$ in the free-energy-density functional (\ref{FPDW}). To compute these terms, we substitute in $f^{(3)}[\phi]$ and $f^{(4)}[\phi]$ the approximate mode expansion $\phi\!\approx \!\sum_\b{q}u_\b{q}\Phi_\b{q}^{(j_*)}(z)e^{i\b{q}\cdot\b{r}_\parallel}$ that discards the noncritical modes $\Phi_\b{q}^{(j)}(z)$ with $j\!\neq\!j_*$. One obtains $f^{(k)}[\phi]\!\approx\!\sum_{\b{q}_1,\ldots,\b{q}_k}\Gamma^{(k)}(\b{q}_1,\ldots,\b{q}_k)u_{\b{q}_1}\cdots u_{\b{q}_k}\delta_{\sum_{j=1}^k{\b{q}_j},0}$, where the $k$-point vertex $\Gamma^{(k)}$ is in general momentum-dependent~\cite{suppmat}, and the corresponding interaction nonlocal in position space. In the spirit of a gradient expansion, we approximate this nonlocal interaction by a local interaction $f^{(k)}[\phi]\!\propto\!\int d^2r_\parallel\,u^k(\b{r}_\parallel)$ obtained by setting to zero all momenta in $\Gamma^{(k)}$. The cubic and quartic terms in (\ref{FPDW}) become:
\begin{align}
    f^{(3)}[\phi]&\approx-w\sum_{\b{q}_1,\b{q}_2,\b{q}_3}u_{\b{q}_1}u_{\b{q}_2}u_{\b{q}_3}\delta_{\b{q}_1+\b{q}_2+\b{q}_3,0},
    \label{f3PDW}\\
    f^{(4)}[\phi]&\approx\eta\sum_{\b{q}_1,\b{q}_2,\b{q}_3,\b{q}_4}u_{\b{q}_1}u_{\b{q}_2}u_{\b{q}_3}u_{\b{q}_4}\delta_{\b{q}_1+\b{q}_2+\b{q}_3+\b{q}_4,0},\label{f4PDW}
\end{align}
where $w\!\propto\!\Gamma^{(3)}(0,\ldots,0)$ and $\eta\!\propto\!\Gamma^{(4)}(0,\ldots,0)$. We find that $\eta\!>\!0$ regardless of temperature, pressure, and confinement~\cite{suppmat}, thus $f_\text{PDW}$ in (\ref{FPDW}) is properly bounded from below.

Equations (\ref{f2PDW}), (\ref{f3PDW}), and (\ref{f4PDW}) define a well-behaved Landau functional for the PDW transition in confined $^3$He, which is the first main result of this Letter. This Landau functional being mathematically identical to that for the crystallization transition in classical statistical mechanics~\cite{landau1937,ChaikinLubensky}, we simply transpose well-known results for the latter to the PDW transition. Restricting our analysis to momenta $\b{q}$ with fixed magnitude $|\b{q}|\!=\!Q$, since at $r\!=\!0$ only modes with such momenta soften, we identity the PDW phase as that in which order-parameter configurations $u_\b{q}\!\neq\!0$ globally minimize $f_\text{PDW}$. Provided $w\!\neq\!0$, the continuous degeneracy of such configurations is partially lifted by the cubic term (\ref{f3PDW}), which favors a PDW with noncollinear wavevectors such that $\b{q}_1\!+\!\b{q}_2\!+\!\b{q}_3\!=\!0$, i.e., that add up to zero in triangles. In the planar phase at $T_2^*$, we find $w\!=\!0$ and the degeneracy remains unresolved at the mean-field level~\cite{suppmat}. In the PDB phase at $T_1^*$ however, we find $w\!\neq\!0$ at all pressures and the noncollinear constraint applies. In 2D, this necessarily implies a PDW whose reciprocal lattice is triangular~\cite{ChaikinLubensky}, corresponding to a triangular Bravais lattice with lattice constant $a=4\pi/(\sqrt{3}Q)$ in position space. This conclusion is the second main result of this Letter.

To be explicit, we set the PDW order parameter $u_\b{q}$ to a constant $u$ for $\b{q}\!\in\!\{\pm\b{G}_1,\pm\b{G}_2,\pm\b{G}_3\}$ and to zero otherwise, where we can take $\b{G}_1\!=\!Q(1,0)$, $\b{G}_2\!=\!Q(-\frac{1}{2},\frac{\sqrt{3}}{2})$, and $\b{G}_3\!=\!-\b{G}_1\!-\!\b{G}_2$ without loss of generality ($f_\text{PDW}$ is invariant under a global rotation). The Landau functional (\ref{FPDW}) then reduces to a simple function of $u$,
\begin{align}
    f_\text{PDW}(u)=6ru^2-12wu^3+216\eta u^4,
\end{align}
whose phase diagram is well understood~\cite{ChaikinLubensky}. Assuming fixed pressure and sample thickness for illustration, we can write $r\!=\!b(T^*_1\!-\!T)$ in the vicinity of the PDW transition from the PDB side, where $b\!>\!0$. Due to the cubic term, the mode softening instability at $T_1^*$ is preempted by a first-order transition at a lower temperature $T_\text{PDW}\!<\!T^*_1$,
\begin{align}\label{TPDW}
    T_\text{PDW}=T^*_1-\frac{w^2}{36\eta b},
\end{align}
which is the true PDW transition temperature. We find the difference $T_1^*\!-\!T_\text{PDW}$ is approximately 4\% of the width $T_2^*\!-\!T_1^*$ of the instability region for a wide range of pressures and sample thicknesses [Fig.~\ref{fig:phasediagram}(b)]. $T^*_1$ is then understood as the limit of metastability of the uniform (PDB) phase inside the PDW phase. Conversely, the PDW phase exists as a metastable phase for $T^{**}\!<\!T\!<\!T_\text{PDW}$ where $T^{**}\!=\!T_\text{PDW}\!-\!w^2/(288\eta b)$.

\begin{figure}[t]
    \includegraphics[width=\columnwidth]{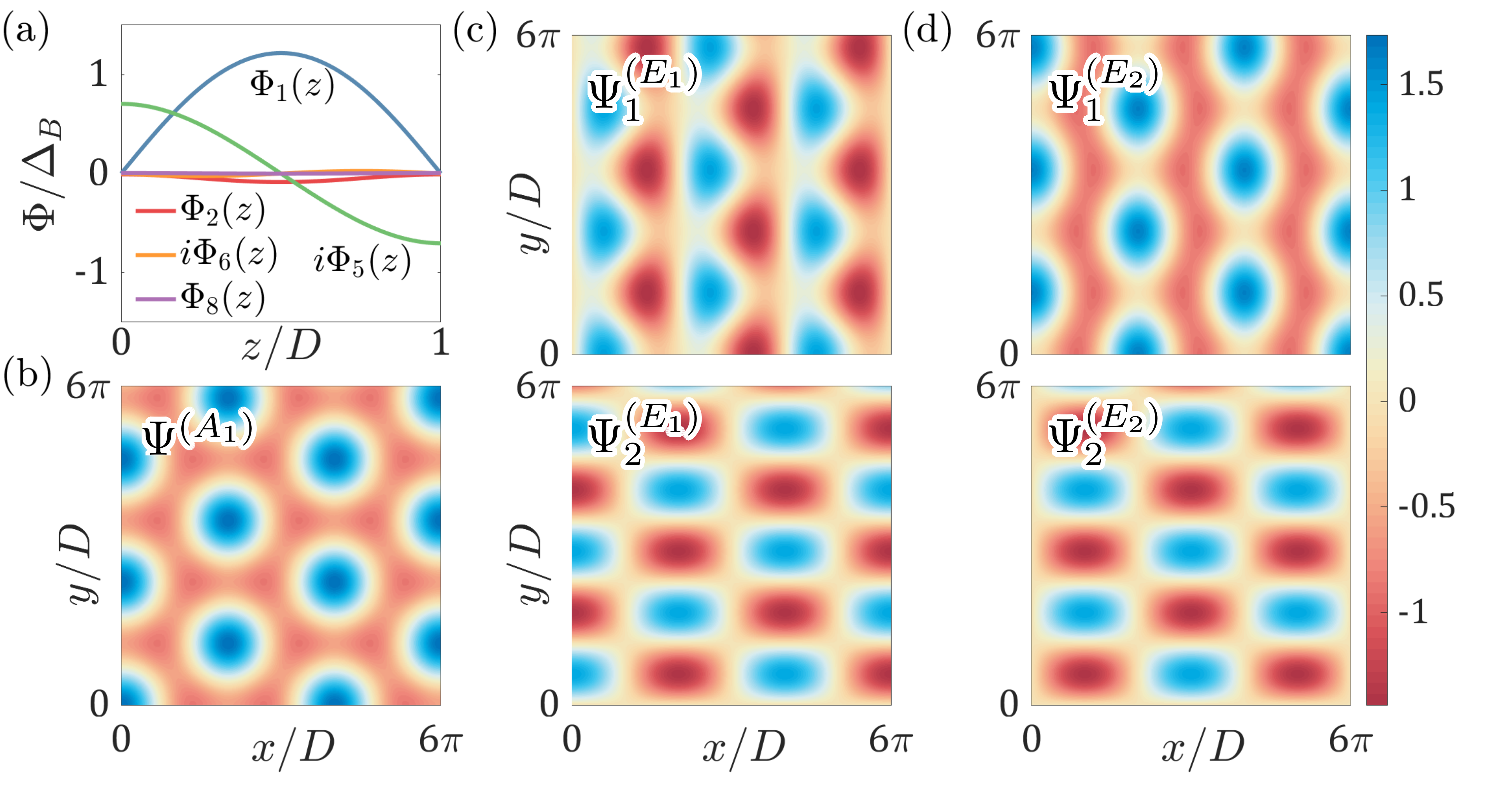}
    \caption{(a) Irreducible $z$-dependent amplitudes of the PDW order parameter; these multiply invariants of the triangular spin-orbital point group that involve $\b{r}_\parallel$-dependent basis functions $\Psi(\b{r}_\parallel)$ for the (b) $A_1$, (c) $E_1$, and (d) $E_2$ irreducible representations of $D_6$.}
    \label{fig:PDW}
    \centering
\end{figure}

{\it Triangular lattice PDW.---}We finally turn to the detailed structure of the PDW order parameter. In the PDW phase, the continuous $SO(2)_{L_z+S_z}$ spin-orbital rotation symmetry is spontaneously broken to a discrete $C_6^{L_z+S_z}$ subgroup of joint spin-orbital rotations. The PDW order parameter $\phi\!=\!u\sum_{\b{q}}\Phi_\b{q}^{(j_*)}(z)e^{i\b{q}\cdot\b{r}_\parallel}$, where the sum is now restricted to the six wavevectors $\b{q}\!\in\!\{\pm\b{G}_1,\pm\b{G}_2,\pm\b{G}_3\}$, is in fact invariant under the full spin-orbital point group $D_6^{L_z+S_z}$,
\begin{align}
    \c{D}(g)\phi(g^{-1}\b{r}_\parallel,z)\c{D}^{-1}(g)=\phi(\b{r}_\parallel,z),\hspace{5mm}g\in D_6,
\end{align}
where the representation matrices $\c{D}(g)$ act on both orbital and spin indices. Using the theory of invariants~\cite{BirPikus}, $\phi$ can be expressed as $\phi\!=\!u\sum_{j,k}\phi^{(j,k)}(z)X^{(j,k)}(\b{r}_\parallel)$ where $X^{(j,k)}(\b{r}_\parallel)$ denotes the $k$th $D_6^{L_z+S_z}$ invariant associated with the irreducible representation $j$ of $D_6$, and $\phi^{(j,k)}(z)$ the corresponding nonuniversal amplitude encapsulating the $z$ dependence of the PDW order parameter~\cite{suppmat}. Only five such amplitudes are nonzero [Fig.~\ref{fig:PDW}(a), also Fig.~(S1) in Ref.~\cite{suppmat}]; the corresponding $X$ invariants involve basis functions for the irreducible $A_1$, $E_1$, $E_2$ representations [Fig.~\ref{fig:PDW}(b,c,d)], to be understood as triangular lattice harmonics with angular momentum $\ell\!=\!0,1,2$, respectively. The $z$ dependence of the amplitudes in Fig.~\ref{fig:PDW}(a) is such that $\phi$ is also invariant under reflection in the $z$ direction: $M_z\phi(\b{r}_\parallel,-z)M_z^{-1}\!=\!\phi(\b{r}_\parallel,z)$, where $M_z\!=\!\diag(1,1,-1)$ acts on both orbital and spin indices.

{\it Conclusion.---}In summary, we have proposed a general mechanism whereby a 2D PDW with hexagonal symmetry is stabilized in confined $^3$He: the Landau functional for a PDW transition {\it within} the uniform superfluid generically contains a cubic term which, upon approach to a crystallization instability, leads to a first-order transition to a PDW with noncollinear wavevectors forming a triangular lattice. We demonstrated that in the weak-coupling approximation, this mechanism is operative for a wide range of sample thicknesses and pressures near the crystallization instability within the PDB phase.

In our weak-coupling treatment, the coefficient of the cubic term was found to vanish at the PDW instability of the planar phase ($T_2^*$). In reality, strong-coupling effects stabilize the A phase over the planar phase in this part of the phase diagram, and Eq.~(\ref{PDB}) is the wrong expansion point. The construction of a PDW Landau functional using the A phase as a starting point and incorporating strong-coupling corrections would be necessary to address the question whether the A-PDW transition remains continuous or also becomes first-order. Additionally, direct numerical minimization of the 3D GL functional using a $D_6^{L_z+S_z}$-invariant PDW ansatz beyond the single-harmonic approximation utilized here would be desirable for more quantitative comparisons with experiment.

{\it Acknowledgements.---}We thank J.P. Davis, F. Marsiglio, L. Radzihovsky, A. Shook, and A. Vorontsov for useful discussions. P.S.Y. was supported by the Alberta Innovates Graduate Student Scholarship Program. R.B. was supported by D\'epartement de physique, Universit\'e de Montr\'eal. J.M. was supported by NSERC Discovery Grants Nos. RGPIN-2014-4608, RGPIN-2020-06999, RGPAS-2020-00064; the CRC Program; CIFAR; a Government of Alberta MIF Grant; a Tri-Agency NFRF Grant (Exploration Stream); and the PIMS CRG program.

\bibliography{3He}

\begin{thebibliography}{44}
\bibitem [{Not()}]{Note2d}%
  \BibitemOpen
  \href@noop {} {}\bibinfo {note} {A 2D PDW is crucially distinguished from the
  inhomogeneous superfluid phase studied in Ref.~\cite{wiman2014} by the fact
  that the breaking of translation symmetry is explicit (external) in the
  latter, but spontaneous in the former.}\BibitemShut {Stop}%
\end{thebibliography}

\end{document}


\newcommand{\tr}{\mathop{\mathrm{tr}}}
\newcommand{\bsigma}{\boldsymbol{\sigma}}
\newcommand{\re}{\mathop{\mathrm{Re}}}
\newcommand{\im}{\mathop{\mathrm{Im}}}
\renewcommand{\b}[1]{{\boldsymbol{#1}}}
\renewcommand{\c}[1]{\mathcal{#1}}
\newcommand{\diag}{\mathrm{diag}}
\newcommand{\sign}{\mathrm{sign}}
\newcommand{\sgn}{\mathop{\mathrm{sgn}}}

\newcommand{\cl}{\mathrm{cl}}
\newcommand{\mb}{\bm}
\newcommand{\ua}{\uparrow}
\newcommand{\da}{\downarrow}
\newcommand{\ra}{\rightarrow}
\newcommand{\la}{\leftarrow}
\newcommand{\mc}{\mathcal}
\newcommand{\bs}{\boldsymbol}
\newcommand{\lra}{\leftrightarrow}
\newcommand{\nn}{\nonumber}
\newcommand{\half}{{\textstyle{\frac{1}{2}}}}
\newcommand{\mf}{\mathfrak}
\newcommand{\MF}{\text{MF}}
\newcommand{\IR}{\text{IR}}
\newcommand{\UV}{\text{UV}}
\newcommand{\sech}{\mathrm{sech}}

\renewcommand{\thetable}{S\Roman{table}}
\renewcommand{\thefigure}{S\arabic{figure}}
\renewcommand{\thesection}{S\Roman{section}}
\renewcommand{\thesubsection}{\Alph{subsection}}
\renewcommand{\theequation}{S\arabic{equation}}

\title{Supplemental Material for ``Triangular pair density wave in confined superfluid $^3$He''}

\author{Pramodh Senarath Yapa}
\affiliation{Department of Physics, University of Alberta, Edmonton, Alberta T6G 2E1, Canada}

\author{Rufus Boyack}
\affiliation{Department of Physics, University of Alberta, Edmonton, Alberta T6G 2E1, Canada}
\affiliation{Theoretical Physics Institute, University of Alberta, Edmonton, Alberta T6G 2E1, Canada}

\author{Joseph Maciejko}
\affiliation{Department of Physics, University of Alberta, Edmonton, Alberta T6G 2E1, Canada}
\affiliation{Theoretical Physics Institute, University of Alberta, Edmonton, Alberta T6G 2E1, Canada}

\date\today

\maketitle

In this Supplemental Material, we explain the decomposition of the quadratic term in the pair density wave (PDW) Landau functional into normal modes (Sec.~\ref{sec:quadratic}), the derivation of the cubic and quartic terms (Sec.~\ref{sec:cubic}), and the decomposition of the PDW order parameter into invariants of its symmetry group, the $D_6^{L_z+S_z}$ spin-orbital point group (Sec.~\ref{sec:symmetry}).

\section{Quadratic term in the PDW functional}\label{sec:quadratic}

The starting point is the Ginzburg-Landau (GL) functional in Eq.~(1) of the main text,
\begin{align}\label{GL}
F=\int d^3r\Bigl[&K_1\partial_k A_{\mu j}\partial_k A_{\mu j}^*
+K_2\partial_j A_{\mu j}\partial_k A_{\mu k}^*
+K_3\partial_k A_{\mu j}\partial_j A_{\mu k}^*
+\alpha \tr AA^\dag\nn\\
&+\beta_1|\tr AA^T|^2+\beta_2(\tr AA^\dag)^2
+\beta_3 \tr AA^T(AA^T)^*+\beta_4\tr(AA^\dag)^2
+\beta_5\tr AA^\dag(AA^\dag)^*\Bigr],
\end{align}
where $\b{r}=(x,y,z)$. We write the order parameter as
\begin{align}\label{OPexpansion}
A_{\mu j}(\b{r}_\parallel,z)=\overline{A}_{\mu j}(z)+\sum_\b{q}\phi_{\mu j}(\b{q},z)e^{i\b{q}\cdot\b{r}_\parallel},
\end{align}
where $\b{r}_\parallel=(x,y)$, $\b{q}=(q_x,q_y)$ and
\begin{align}\label{PDBOP}
\overline{A}_{\mu j}(z)=\left(\begin{array}{ccc}
\Delta_\parallel(z) & 0 & 0 \\
0 & \Delta_\parallel(z) & 0 \\
0 & 0 & \Delta_\perp(z)
\end{array}\right),
\end{align}
is the equilibrium order parameter in the planar-distorted B phase (PDB phase). We assume $\phi_{\mu j}\ll \Delta_\parallel,\Delta_\perp$ and wish to expand the GL functional to quartic order in $\phi$. The leading nontrivial term in this expansion is the term quadratic in $\phi$, which will be of the form
\begin{align}\label{FreeEnergy}
f=f[\overline{A}]+\sum_\b{q}\int_0^D dz\,\phi_{\mu j}^*(\b{q},z)\hat{C}_{\mu j,\nu k}(\b{q},z)\phi_{\nu k}(\b{q},z)+\mathcal{O}(\phi^3),
\end{align}
by translational invariance, where $f$ denotes the free energy per unit area of the film. By diagonalizing $\hat{C}$ we will be able to identity the soft mode which is relevant for the crystallization transition, as done in Ref.~\cite{mizushima2018}. We use the hatted notation to indicate that $\hat{C}$ is an operator acting on the $z$ dependence of functions to its right. Also note that since $\overline{A}$ is a stationary point of $F$, one has $\delta F/\delta{\overline{A}}=\delta F/\delta\overline{A}^*=0$ and there are no terms linear in $\phi$. To diagonalize $\hat{C}_{\mu j,\nu k}(\b{q})$, it is simplest to proceed as in $\b{k}\cdot\b{p}$ theory for the calculation of electronic bandstructures (see, e.g., Ref.~\cite{Winkler}): we will use the (``band-edge'') eigenvectors of $\hat{C}_{\mu j,\nu k}(\b{0})$ as a basis for the eigenfunctions of $\hat{C}_{\mu j,\nu k}(\b{q})$.

\subsection{Quadratic term: \texorpdfstring{$\boldsymbol{q}=0$}{G0}}\label{sec:q0}

We therefore start by assuming $\b{q}=0$, i.e., $A_{\mu j}(\b{r}_\parallel,z)=\overline{A}_{\mu j}(z)+\phi_{\mu j}(z)$. Since there is no $\b{r}_\parallel$ dependence, only gradient terms involving $\partial_z$ survive in the GL functional. First, the GL functional is invariant under the time-reversal transformation
\begin{align}
\mathcal{T}\colon A_{\mu j}\rightarrow A_{\mu j}^*.
\end{align}
Assuming $\Delta_\parallel$ and $\Delta_\perp$ real, the PDB order parameter (\ref{PDBOP}) is invariant under $\mathcal{T}$. To quadratic order, real and imaginary fluctuations cannot mix~\cite{VW3He}, thus we write $\phi_{\mu j}=\phi_{\mu j}^{(+)}+i\phi_{\mu j}^{(-)}$ with $\phi_{\mu j}^{(\pm)}$ real. Each $3\times 3$ matrix has 9 real components and can be decomposed as the sum of a trace part (proportional to the identity) and a traceless real matrix with 8 components. The latter can be decomposed into symmetric traceless (5 components) and antisymmetric (3 components) parts. A useful representation is in terms of the $SU(3)$ Gell-Mann matrices $\lambda^1,\ldots,\lambda^8$ augmented with a matrix $\lambda^0$ proportional to the identity,
\begin{align}\label{GellMann}
\lambda^0&=\sqrt{\frac{2}{3}}\left(\begin{array}{ccc}
1 & & \\
& 1 & \\
& & 1
\end{array}\right),\nn\\
\lambda^1&=\left(\begin{array}{ccc}
1 & & \\
& -1 & \\
& & 0
\end{array}\right),\hspace{2mm}
\lambda^2=\frac{1}{\sqrt{3}}\left(\begin{array}{ccc}
1 & & \\
& 1 & \\
& & -2
\end{array}\right),\hspace{2mm}
\lambda^3=\left(\begin{array}{ccc}
& &  \\
& & 1 \\
& 1 & 
\end{array}\right),\hspace{2mm}
\lambda^4=\left(\begin{array}{ccc}
& & 1 \\
& &  \\
1 &  & 
\end{array}\right),\hspace{2mm}
\lambda^5=\left(\begin{array}{ccc}
& 1 &  \\
1 & &  \\
&  & 
\end{array}\right),\nn\\
\lambda^6&=\left(\begin{array}{ccc}
& &  \\
& & -i \\
& i & 
\end{array}\right)=L_x,\hspace{2mm}
\lambda^7=\left(\begin{array}{ccc}
&  & i  \\
 & &  \\
-i &  & 
\end{array}\right)=L_y,\hspace{2mm}
\lambda^8=\left(\begin{array}{ccc}
& -i &  \\
i & &  \\
&  & 
\end{array}\right)=L_z,
\end{align}
and are normalized according to $\tr \lambda^I\lambda^J=2\delta^{IJ},\,I,J=0,\ldots,8$. We have also observed that $(\lambda^6,\lambda^7,\lambda^8)=(L_x,L_y,L_z)=\b{L}$ are the $J=1$ angular momentum matrices, i.e., the generators of $SO(3)$ in the vector representation. They are purely imaginary, thus we can expand
\begin{align}\label{phiExpansion}
\phi^{(\pm)}=\Delta^{(\pm)}\lambda^0+i\b{\Omega}^{(\pm)}\cdot\b{L}+\b{M}^{(\pm)}\cdot\b{\Lambda},
\end{align}
where $\Delta^{(\pm)}$ is the scalar ($J=0$) mode, $\b{\Omega}^{(\pm)}=(\Omega_1^{(\pm)},\Omega_2^{(\pm)},\Omega_3^{(\pm)})$ is the vector ($J=1$) mode, and $\b{M}^{(\pm)}=(M_1^{(\pm)},\ldots,M_5^{(\pm)})$ is the symmetric traceless tensor ($J=2$) mode, defining $\b{\Lambda}=(\lambda^1,\ldots,\lambda^5)$.

We substitute the expansion (\ref{phiExpansion}) into the GL functional and expand to quadratic order in $\phi$. We ignore the equilibrium free energy $f[\overline{A}]$, i.e., we consider the free energy relative to that of the equilibrium PDB phase. The quadratic term in the GL functional gives
\begin{align}\label{quadratic1}
f^{(2)}[\phi]=\sum_{K=\pm}\int_0^D dz\biggl\{&2K_1\left[(\partial_z\Delta^{(K)})^2+(\partial_z\b{\Omega}^{(K)})^2+(\partial_z\b{M}^{(K)})^2\right]\nn\\
&+K_{23}\left[\frac{2}{3}(\partial_z\Delta^{(K)}-\sqrt{2}\partial_z M_2^{(K)})^2+(\partial_z M_3^{(K)}+\partial_z\Omega_1^{(K)})^2+(\partial_z M_4^{(K)}-\partial_z\Omega_2^{(K)})^2\right]\nn\\
&+2\alpha\left[(\Delta^{(K)})^2+(\b{\Omega}^{(K)})^2+(\b{M}^{(K)})^2\right]\biggr\},
\end{align}
where $K_{ijk\cdots}=K_i+K_j+K_k+\cdots$. The quartic term gives
\begin{align}\label{quadratic2}
f^{(4)}[\phi]=4\sum_{K=\pm}\int_0^Ddz\biggl\{&c_\Delta^{(K)}(z)(\Delta^{(K)})^2+c_{\Omega,\parallel}^{(K)}(z)\left[(\Omega_1^{(K)})^2+(\Omega_2^{(K)})^2\right]
+c_{\Omega,\perp}^{(K)}(z)(\Omega_3^{(K)})^2+c_{M,m_J=0}^{(K)}(z)(M_2^{(K)})^2\nn\\
&+c_{M,m_J=\pm 1}^{(K)}(z)\left[(M_4^{(K)})^2+(M_3^{(K)})^2\right]
+c_{M,m_J=\pm 2}^{(K)}(z)\left[(M_1^{(K)})^2+(M_5^{(K)})^2\right]\nn\\
&+c_{M\Delta}^{(K)}(z)M_2^{(K)}\Delta^{(K)}
+c_{M\Omega}^{(K)}(z)\left(M_4^{(K)}\Omega_2^{(K)}-M_3^{(K)}\Omega_1^{(K)}\right)\biggr\},
\end{align}
where the $c$ coefficients are functions of the equilibrium order parameters $\Delta_\parallel(z)$, $\Delta_\perp(z)$ and are given by
\begin{align}
c_\Delta^{(+)}(z)&=\frac{1}{3}\left[2(7\beta_{12}+3\beta_{345})\Delta_\parallel^2+8\beta_{12}\Delta_\parallel\Delta_\perp+(5\beta_{12}+3\beta_{345})\Delta_\perp^2\right],\\
c_\Delta^{(-)}(z)&=\frac{1}{3}\left[2(\beta_1+3\beta_2+\beta_{345})\Delta_\parallel^2+8\beta_1\Delta_\parallel\Delta_\perp-(\beta_1-3\beta_2-\beta_{345})\Delta_\perp^2\right],
\end{align}
for terms involving the $J=0$ modes $\Delta^{(\pm)}$ alone,
\begin{align}
c_{\Omega,\parallel}^{(+)}(z)&=(2\beta_{12}+\beta_{345})\Delta_\parallel^2-\beta_{345}\Delta_\parallel\Delta_\perp+\beta_{12345}\Delta_\perp^2,\\
c_{\Omega,\parallel}^{(-)}(z)&=-(2\beta_1-2\beta_2-\beta_4)\Delta_\parallel^2-(\beta_{35}-\beta_4)\Delta_\parallel\Delta_\perp-(\beta_1-\beta_2-\beta_4)\Delta_\perp^2,\\
c_{\Omega,\perp}^{(+)}(z)&=(2\beta_{12}+\beta_{345})\Delta_\parallel^2+\beta_{12}\Delta_\perp^2,\\
c_{\Omega,\perp}^{(-)}(z)&=-(2\beta_1-2\beta_2+\beta_{35}-3\beta_4)\Delta_\parallel^2-(\beta_1-\beta_2)\Delta_\perp^2,
\end{align}
for terms involving the $J=1$ modes $\b{\Omega}^{(\pm)}$ alone,
\begin{align}
c_{M,m_J=0}^{(+)}(z)&=\frac{1}{3}\left[(10\beta_{12}+3\beta_{345})\Delta_\parallel^2-8\beta_{12}\Delta_\parallel\Delta_\perp+(7\beta_{12}+6\beta_{345})\Delta_\perp^2\right],\\
c_{M,m_J=0}^{(-)}(z)&=-\frac{1}{3}\left[(2\beta_1-6\beta_2-\beta_{345})\Delta_\parallel^2+8\beta_1\Delta_\parallel\Delta_\perp-(\beta_1+3\beta_2+2\beta_{345})\Delta_\perp^2\right],\\
c_{M,m_J=\pm 1}^{(+)}(z)&=(2\beta_{12}+\beta_{345})\Delta_\parallel^2+\beta_{345}\Delta_\parallel\Delta_\perp+\beta_{12345}\Delta_\perp^2,\\
c_{M,m_J=\pm 1}^{(-)}(z)&=-(2\beta_1-2\beta_2-\beta_4)\Delta_\parallel^2-(\beta_4-\beta_{35})\Delta_\parallel\Delta_\perp-(\beta_1-\beta_{24})\Delta_\perp^2,\\
c_{M,m_J=\pm 2}^{(+)}(z)
&=(2\beta_{12}+3\beta_{345})\Delta_\parallel^2+\beta_{12}\Delta_\perp^2,\\
c_{M,m_J=\pm 2}^{(-)}(z)&=-(2\beta_1-2\beta_2-\beta_{345})\Delta_\parallel^2-(\beta_1-\beta_2)\Delta_\perp^2,
\end{align}
for terms involving the $J=2$ modes $\b{M}^{(\pm)}$ alone, and
\begin{align}
c_{M\Delta}^{(+)}(z)&=\frac{2\sqrt{2}}{3}(\Delta_\parallel-\Delta_\perp)\left[(4\beta_{12}+3\beta_{345})\Delta_\parallel+(2\beta_{12}+3\beta_{345})\Delta_\perp\right],\\
c_{M\Delta}^{(-)}(z)&=\frac{2\sqrt{2}}{3}(\Delta_\parallel-\Delta_\perp)\left[(4\beta_1+\beta_{345})\Delta_\parallel+(2\beta_1+\beta_{345})\Delta_\perp\right],\\
c_{M\Omega}^{(+)}(z)&=0,\\
c_{M\Omega}^{(-)}(z)&=2(\beta_3-\beta_5)(\Delta_\parallel^2-\Delta_\perp^2),
\end{align}
for couplings between modes. We use the notation $\beta_{ijk\cdots}=\beta_i+\beta_j+\beta_k+\cdots$.

The reduced functional (\ref{quadratic1}-\ref{quadratic2}) should preserve the $SO(2)_{L_z+S_z}$ symmetry of the PDB phase. To determine how $\phi$ transforms under this symmetry, first recall the action of the residual $SO(3)_{L+S}$ symmetry in the bulk B phase~\cite{VW3He}:
\begin{align}\label{SOrotation}
A_{\mu j}\rightarrow R_{\mu\nu}^{(S)}R_{jk}^{(L)}A_{\nu k}=\left(R^{(S)}AR^{(L)T}\right)_{\mu j},
\end{align}
where $A=e^{i\phi}R$, $R^{(L)}=O$, and $R^{(S)}=ROR^{-1}$, with $R,O\in SO(3)$. By relative spin-orbit symmetry, one can choose $R=\mathbb{I}$, in which case the residual symmetry acts by conjugation, $A\rightarrow OAO^{-1}$. For a $J_z$ rotation about the $z$ axis through angle $\theta$, one has $O=e^{-i\theta L_z}$ with $L_z=\lambda^8$ defined earlier. This is clearly a symmetry of the PDB phase since $e^{-i\theta L_z}\overline{A}e^{i\theta L_z}=\overline{A}$. Fluctuations will transform as $\phi\rightarrow e^{-i\theta L_z}\phi e^{i\theta L_z}$; it is sufficient to consider infinitesimal rotations, $\phi\rightarrow \phi+\delta\phi$ where $\delta\phi=-i\theta[L_z,\phi]$. Since $L_z$ is purely imaginary, this transformation does not mix real and imaginary modes (it commutes with time reversal, as expected). One can thus write the first variation as $\delta\phi=\delta\phi^{(+)}+i\delta\phi^{(-)}$ where $\delta\phi^{(\pm)}$ are real, and can thus be expanded as in (\ref{phiExpansion}):
\begin{align}
\delta\phi^{(\pm)}=\delta\Delta^{(\pm)}\lambda^0+i\delta\b{\Omega}^{(\pm)}\cdot\b{L}+\delta\b{M}^{(\pm)}\cdot\b{\Lambda}.
\end{align}
On the other hand, we have
\begin{align}
\delta\phi^{(\pm)}=-i\theta[L_z,\phi^{(\pm)}]
=-i\theta\epsilon_{ij}\Omega_j^{(\pm)}L_i-i\theta M_a^{(\pm)}[\lambda^8,\lambda^a],
\end{align}
where $i,j=1,2$ and $a=1,\ldots,5$, and we have used the commutation relations for the angular momentum operators. We thus find
\begin{align}\label{JzTrafo1}
\delta\Delta^{(\pm)}=0;\hspace{5mm}
\delta\Omega_3^{(\pm)}=0;\hspace{5mm}
\delta\Omega_i^{(\pm)}=-\theta\epsilon_{ij}\Omega_j^{(\pm)},\,i,j=1,2.
\end{align}
This is as expected: $\Delta^{(\pm)}$ is a scalar mode that does not transform under rotations; $\Omega_3^{(\pm)}$ is the $z$ component of a vector mode, that is left invariant under $J_z$ rotations; and $\Omega_1^{(\pm)}$ and $\Omega_2^{(\pm)}$ are the $x$ and $y$ components of this vector mode, which transform into each other under $J_z$ rotations.

Turning to the $J=2$ modes, one can check that $i[\lambda^8,\lambda^a]$ is real symmetric traceless, thus it can be expanded in the basis of $\lambda^a$ matrices:
\begin{align}
-i[\lambda^8,\lambda^a]=\sum_{b=1}^5\mathcal{J}_{ab}\lambda^b,\,a=1,\ldots,5,
\end{align}
with $\mathcal{J}$ a real antisymmetric $5\times 5$ matrix. Using the orthogonality property of Gell-Mann matrices, we have \begin{align}
\mathcal{J}_{ab}=\frac{1}{2i}\tr([\lambda^8,\lambda^a]\lambda^b)=\left(\begin{array}{ccccc}
0 & 0 & 0 & 0 & 2 \\
0 & 0 & 0 & 0 & 0 \\
0 & 0 & 0 & -1 & 0 \\
0 & 0 & 1 & 0 & 0 \\
-2 & 0 & 0 & 0 & 0
\end{array}\right)_{ab}.
\end{align}
Thus we find $\delta M_a^{(\pm)}=-\theta\mathcal{J}_{ab}M_b^{(\pm)}$. More specifically, we have
\begin{align}\label{JzTrafo2}
\delta M_2^{(\pm)}&=0;\nn\\
\delta M_4^{(\pm)}&=-\theta M_3^{(\pm)},\hspace{5mm}
\delta M_3^{(\pm)}=\theta M_4^{(\pm)};\nn\\
\delta M_1^{(\pm)}&=-2\theta M_5^{(\pm)},\hspace{5mm}
\delta M_5^{(\pm)}=2\theta M_1^{(\pm)}.
\end{align}
Once again, there is a clear interpretation: $M_2^{(\pm)}$ is an $m_J=0$ mode (like a $d_{3z^2-r^2}$ orbital), invariant under $J_z$ rotations; $M_4^{(\pm)}$ and $M_3^{(\pm)}$ can be combined to form $m_J=\pm 1$ modes (like $d_{xz}\pm i d_{yz}$ orbitals); and $M_1^{(\pm)}$ and $M_5^{(\pm)}$ can be combined to form $m_J=\pm 2$ modes (like $d_{x^2-y^2}\pm i d_{xy}$ orbitals). Using the transformation properties above, one can check explicitly that the functional (\ref{quadratic1}-\ref{quadratic2}) is invariant under infinitesimal $SO(2)_{J_z}$ rotations.

In order to find the normal modes, we will first bring the gradient terms in Eq.~(\ref{quadratic1}) to diagonal form by performing an orthogonal transformation. In terms of the rotated fields
\begin{align}\label{RotatedFields1}
\tilde{\Delta}^{(K)}=\frac{1}{\sqrt{3}}(\Delta^{(K)}-\sqrt{2}M_2^{(K)}),\hspace{5mm}
\tilde{\b\Omega}^{(K)}=\left(\begin{array}{c}
\tilde{\Omega}_1^{(K)} \\
\tilde{\Omega}_2^{(K)} \\
\tilde{\Omega}_3^{(K)}
\end{array}\right)=
\left(\begin{array}{c}
\frac{1}{\sqrt{2}}(\Omega_1^{(K)}-M_3^{(K)}) \\
\frac{1}{\sqrt{2}}(\Omega_2^{(K)}+M_4^{(K)}) \\
\Omega_3^{(K)}
\end{array}\right),
\end{align}
and
\begin{align}\label{RotatedFields2}
\tilde{\b{M}}^{(K)}=\left(\begin{array}{c}
\tilde{M}_1^{(K)} \\
\tilde{M}_2^{(K)} \\
\tilde{M}_3^{(K)} \\
\tilde{M}_4^{(K)} \\
\tilde{M}_5^{(K)}
\end{array}\right)
=\left(\begin{array}{c}
M_1^{(K)} \\
\frac{1}{\sqrt{3}}(M_2^{(K)}+\sqrt{2}\Delta^{(K)}) \\
\frac{1}{\sqrt{2}}(M_3^{(K)}+\Omega_1^{(K)}) \\
\frac{1}{\sqrt{2}}(M_4^{(K)}-\Omega_2^{(K)}) \\
M_5^{(K)}
\end{array}\right),
\end{align}
Equation~(\ref{quadratic1}) becomes
\begin{align}
f^{(2)}[\phi]=\sum_{K=\pm}\int_0^Ddz\Bigl\{&2K_{123}\left[(\partial_z\tilde{\Delta}^{(K)})^2
+(\partial_z\tilde{M}_3^{(K)})^2+(\partial_z\tilde{M}_4^{(K)})^2\right]\nn\\
&+2K_1\left[(\partial_z\tilde{\b{\Omega}}^{(K)})^2+(\partial_z\tilde{M}_2^{(K)})^2+(\partial_z\tilde M_1^{(K)})^2+(\partial_z\tilde M_5^{(K)})^2\right]\nn\\
&+2\alpha\left[(\tilde{\Delta}^{(K)})^2+(\tilde{\b{\Omega}}^{(K)})^2+(\tilde{\b{M}}^{(K)})^2\right]\Bigr\},
\end{align}
and Eq.~(\ref{quadratic2}) assumes the same form but with different (tilded) coefficients:
\begin{align}
f^{(4)}[\phi]=4\sum_{K=\pm}\int_0^Ddz\biggl\{&\tilde c_\Delta^{(K)}(z)(\tilde\Delta^{(K)})^2+\tilde c_{\Omega,\parallel}^{(K)}(z)\left[(\tilde\Omega_1^{(K)})^2+(\tilde\Omega_2^{(K)})^2\right]
+\tilde c_{\Omega,\perp}^{(K)}(z)(\tilde\Omega_3^{(K)})^2+\tilde c_{M,m_J=0}^{(K)}(z)(\tilde M_2^{(K)})^2\nn\\
&+\tilde c_{M,m_J=\pm 1}^{(K)}(z)\left[(\tilde M_4^{(K)})^2+(\tilde M_3^{(K)})^2\right]
+\tilde c_{M,m_J=\pm 2}^{(K)}(z)\left[(\tilde M_1^{(K)})^2+(\tilde M_5^{(K)})^2\right]\nn\\
&+\tilde c_{M\Delta}^{(K)}(z)\tilde M_2^{(K)}\tilde\Delta^{(K)}
+\tilde c_{M\Omega}^{(K)}(z)\left(\tilde M_4^{(K)}\tilde\Omega_2^{(K)}-\tilde M_3^{(K)}\tilde\Omega_1^{(K)}\right)\biggr\},
\end{align}
with
\begin{align}
\tilde c_\Delta^{(+)}(z)&=2\beta_{12}\Delta_\parallel^2+3(\beta_{12}+\beta_{345})\Delta_\perp^2,\\
\tilde c_\Delta^{(-)}(z)&=-2(\beta_1-\beta_2)\Delta_\parallel^2+\beta_{12345}\Delta_\perp^2,
\end{align}

\begin{align}
\tilde c_{\Omega,\parallel}^{(+)}(z)&=(2\beta_{12}+\beta_{345})\Delta_\parallel^2+(\beta_{12}+\beta_{345})\Delta_\perp^2,\\
\tilde c_{\Omega,\parallel}^{(-)}(z)&=-(2\beta_1-2\beta_2-\beta_3-\beta_4+\beta_5)\Delta_\parallel^2-(\beta_1-\beta_2+\beta_3-\beta_4-\beta_5)\Delta_\perp^2,\\
\tilde c_{\Omega,\perp}^{(+)}(z)&=(2\beta_{12}+\beta_{345})\Delta_\parallel^2+\beta_{12}\Delta_\perp^2,\\
\tilde c_{\Omega,\perp}^{(-)}(z)&=-(2\beta_1-2\beta_2+\beta_3-3\beta_4+\beta_5)\Delta_\parallel^2-(\beta_1-\beta_2)\Delta_\perp^2,
\end{align}

\begin{align}
\tilde c_{M,m_J=0}^{(+)}(z)&=3(2\beta_{12}+\beta_{345})\Delta_\parallel^2+\beta_{12}\Delta_\perp^2,\\
\tilde c_{M,m_J=0}^{(-)}(z)&=(2\beta_{12}+\beta_{345})\Delta_\parallel^2-(\beta_1-\beta_2)\Delta_\perp^2,\\
\tilde c_{M,m_J=\pm 1}^{(+)}(z)&=\tilde c_{\Omega,\parallel}^{(+)}(z)=(2\beta_{12}+\beta_{345})\Delta_\parallel^2+(\beta_{12}+\beta_{345})\Delta_\perp^2,\\
\tilde c_{M,m_J=\pm 1}^{(-)}(z)&=-(2\beta_1-2\beta_2+\beta_3-\beta_4-\beta_5)\Delta_\parallel^2-(\beta_1-\beta_2-\beta_3-\beta_4+\beta_5)\Delta_\perp^2,\\
\tilde c_{M,m_J=\pm 2}^{(+)}(z)&=(2\beta_{12}+3\beta_{345})\Delta_\parallel^2+\beta_{12}\Delta_\perp^2,\\
\tilde c_{M,m_J=\pm 2}^{(-)}(z)&=-(2\beta_1-2\beta_2-\beta_{345})\Delta_\parallel^2-(\beta_1-\beta_2)\Delta_\perp^2,
\end{align}

\begin{align}
\tilde c_{M\Delta}^{(+)}(z)&=4\sqrt{2}\beta_{12}\Delta_\parallel\Delta_\perp,\\
\tilde c_{M\Delta}^{(-)}(z)&=4\sqrt{2}\beta_1\Delta_\parallel\Delta_\perp,\\
\tilde c_{M\Omega}^{(+)}(z)&=2\beta_{345}\Delta_\parallel\Delta_\perp,\\
\tilde c_{M\Omega}^{(-)}(z)&=2(\beta_3-\beta_4+\beta_5)\Delta_\parallel\Delta_\perp.
\end{align}

Using the definitions (\ref{RotatedFields1}-\ref{RotatedFields2}) and the transformation properties (\ref{JzTrafo1}) and (\ref{JzTrafo2}), we can show that the tilded fields transform under $SO(2)_{J_z}$ rotations in the same way as the old ones, i.e.,
\begin{align}
\delta\tilde{\Delta}^{(K)}&=\delta\tilde{\Omega}_3^{(K)}=\delta\tilde{M}_2^{(K)}=0,\\
\delta\tilde\Omega_1^{(K)}&=-\theta\tilde\Omega_2^{(K)},\hspace{5mm}\delta\tilde\Omega_2^{(K)}=\theta\tilde\Omega_1^{(K)},\\
\delta\tilde M_4^{(K)}&=-\theta\tilde M_3^{(K)},\hspace{5mm}\delta\tilde M_3^{(K)}=\theta\tilde M_4^{(K)},\\
\delta\tilde M_1^{(K)}&=-2\theta\tilde M_5^{(K)},\hspace{5mm}\delta\tilde M_5^{(K)}=2\theta\tilde M_1^{(K)}.
\end{align}

The functional can be written in block-diagonal form. Schematically, we have:
\begin{align}
\{\tilde\Delta,\tilde M_2\}\oplus\{\tilde\Omega_1,\tilde M_3\}\oplus\{\tilde\Omega_2,\tilde M_4\}\oplus\tilde\Omega_3\oplus\tilde M_1\oplus\tilde M_5.
\end{align}
We begin with the $J=0,2$, $m_J=0$ block $\{\tilde\Delta,\tilde M_2\}$. We consider specular boundary conditions at $z=0$ and $z=D$, and the tilded quantities obey the same (Neumann or Dirichlet) boundary conditions as $\overline{A}$. Integration by parts thus gives
\begin{align}
\int_0^Ddz\,(\partial_z \tilde{g})^2=-\int_0^D dz\,\tilde{g}\partial_z^2 \tilde{g},
\end{align}
for a function $\tilde{g}$ satisfying such boundary conditions. Using the notation $f_{J;m_J}$ for each block's contribution to the GL functional and $\hat\Pi^{(K)}_{J;m_J}$ for the corresponding quadratic kernel, we have
\begin{align}
f_{0,2;0}=\sum_{K=\pm}\int_0^D dz\,\left(\begin{array}{c}
\tilde{\Delta}^{(K)} \\
\tilde{M}_2^{(K)}
\end{array}\right)^T
\hat\Pi_{0,2;0}^{(K)}
\left(\begin{array}{c}
\tilde{\Delta}^{(K)} \\
\tilde{M}_2^{(K)}
\end{array}\right),
\end{align}
where
\begin{align}\label{PiJ020}
\left(\hat\Pi_{0,2;0}^{(K)}\right)_{11}&=-2K_{123}\partial_z^2+2\alpha+4\tilde{c}_{\Delta}^{(K)}(z),\nn\\
\left(\hat\Pi_{0,2;0}^{(K)}\right)_{12}&=\left(\hat\Pi_{0,2;0}^{(K)}\right)_{21}=2\tilde{c}_{M\Delta}^{(K)}(z),\nn\\
\left(\hat\Pi_{0,2;0}^{(K)}\right)_{22}&=-2K_1\partial_z^2+2\alpha+4\tilde{c}_{M,m_J=0}^{(K)}(z).
\end{align}
Using these expressions we can identify the $U(1)$ (phase) Goldstone mode. This corresponds to pure phase fluctuations of the order parameter:
\begin{align}
A(\b{r})=\overline{A}(z)e^{i\varphi(\b{r})}\approx\overline{A}(z)(1+i\varphi),
\end{align}
thus $\phi_{U(1)}=i\overline{A}\varphi$, implying $\phi^{(+)}_{U(1)}=0$ and $\phi^{(-)}_{U(1)}=\overline{A}\varphi$. Goldstone's theorem implies that uniform fluctuations, i.e., $\varphi=\text{const.}$, cost zero energy. To verify this, we need to show that $\hat{\Pi}_{0,2;0}^{(-)}$ acting on a fluctuation of this type gives zero. Decomposing $\phi^{(-)}_{U(1)}$ as in Eq.~(\ref{phiExpansion}), and using Eqs.~(\ref{RotatedFields1}-\ref{RotatedFields2}) to express the result in terms of the tilded fields, we find
\begin{align}
\tilde{\Delta}_{U(1)}^{(-)}=\frac{\varphi}{\sqrt{2}}\Delta_\perp,\hspace{5mm}
\tilde{M}_{2,U(1)}^{(-)}=\varphi\Delta_\parallel.
\end{align}
(Since $\phi^{(-)}_{U(1)}$ is diagonal, only $\lambda^0$, $\lambda^1$, and $\lambda^2$ may contribute; since $\overline{A}_{xx}=\overline{A}_{yy}$, there is in fact no $\lambda^1$ component.) Using Eq.~(\ref{PiJ020}), we find
\begin{align}\label{GoldstoneCheckJ020}
\hat\Pi_{0,2;0}^{(-)}
\left(\begin{array}{c}
\tilde{\Delta}^{(-)}_{U(1)} \\
\tilde{M}_{2,U(1)}^{(-)}
\end{array}\right)=\sqrt{2}\varphi\left(
\begin{array}{c}
\left[-K_{123}\partial_z^2+\alpha+2\beta_{12345}\Delta_\perp^2+4\beta_{12}\Delta_\parallel^2\right]\Delta_\perp \\
\sqrt{2}\left[-K_1\partial_z^2+\alpha+2(2\beta_{12}+\beta_{345})\Delta_\parallel^2+2\beta_{12}\Delta_\perp^2\right]\Delta_\parallel
\end{array}\right).
\end{align}
On the other hand, one verifies that the GL equations $\delta f/\delta\overline{A}=0$ for $\Delta_\parallel(z)$ and $\Delta_\perp(z)$ are precisely
\begin{align}
\left[-K_1\partial_z^2+\alpha+2(2\beta_{12}+\beta_{345})\Delta_\parallel^2+2\beta_{12}\Delta_\perp^2\right]\Delta_\parallel&=0,\\
\left[-K_{123}\partial_z^2+\alpha+2\beta_{12345}\Delta_\perp^2+4\beta_{12}\Delta_\parallel^2\right]\Delta_\perp&=0,
\end{align}
thus the right-hand side of Eq.~(\ref{GoldstoneCheckJ020}) identically vanishes, provided that the order parameter profiles $\Delta_\parallel(z)$ and $\Delta_\perp(z)$ satisfy the exact GL equations.

Besides the $U(1)$ Goldstone mode, from the symmetry-breaking pattern $SO(3)_S\times SO(2)_{L_z}\rightarrow SO(2)_{L_z+S_z}$ for the PDB phase in a slab geometry we expect 3 more Goldstone modes, 2 corresponding to ordinary spin waves (broken $S_x$ and $S_y)$ rotations and 1 corresponding to a spin-orbit mode (broken relative $L_z-S_z$ rotations). Beginning with the spin waves, they correspond to pure spin rotations about the $\hat{\b{x}}$ and $\hat{\b{y}}$ axes:
\begin{align}
A_{\mu j}(\b{r})=R^{(S)}_{\mu\nu}(\hat{\b{x}},\varphi(\b{r}))\overline{A}_{\nu j}(z)\hspace{5mm}
\text{or}\hspace{5mm}
A_{\mu j}(\b{r})=R^{(S)}_{\mu\nu}(\hat{\b{y}},\varphi(\b{r}))\overline{A}_{\nu j}(z),
\end{align}
with $R^{(S)}(\hat{\b{x}},\varphi)=e^{-i\varphi L_x}$ and $R^{(S)}(\hat{\b{y}},\varphi)=e^{-i\varphi L_y}$. Expanding for small $\varphi$, we find that these modes are real with $\phi_{S_x}^{(+)}=-i\varphi L_x\overline{A}$ and $\phi_{S_y}^{(+)}=-i\varphi L_y\overline{A}$. Decomposing in the Gell-Mann basis and rotating the fields, we find that only the $m_J=\pm 1$ modes $\tilde{\Omega}_{1,2}$ and $\tilde{M}_{3,4}$ are involved:
\begin{align}
\tilde{\Omega}_{1,S_x}^{(+)}&=-\frac{\varphi}{\sqrt{2}}\Delta_\parallel,\hspace{5mm}
\tilde{M}_{3,S_x}^{(+)}=-\frac{\varphi}{\sqrt{2}}\Delta_\perp;\\
\tilde{\Omega}_{2,S_y}^{(+)}&=-\frac{\varphi}{\sqrt{2}}\Delta_\parallel,\hspace{5mm}
\tilde{M}_{4,S_y}^{(+)}=\frac{\varphi}{\sqrt{2}}\Delta_\perp,
\end{align}
with all other components zero. The $m_J=\pm 1$ fluctuation free energy is
\begin{align}
f_{1,2;\pm 1}=\sum_{K=\pm}\int_0^D dz\left[\left(\begin{array}{c}
\tilde{\Omega}_1^{(K)} \\
\tilde{M}_3^{(K)}
\end{array}\right)^T
\hat\Pi_{1,2;x}^{(K)}
\left(\begin{array}{c}
\tilde{\Omega}_1^{(K)} \\
\tilde{M}_3^{(K)}
\end{array}\right)
+
\left(\begin{array}{c}
\tilde{\Omega}_2^{(K)} \\
\tilde{M}_4^{(K)}
\end{array}\right)^T
\hat\Pi_{1,2;y}^{(K)}
\left(\begin{array}{c}
\tilde{\Omega}_2^{(K)} \\
\tilde{M}_4^{(K)}
\end{array}\right)\right],
\end{align}
where
\begin{align}
\left(\hat\Pi_{1,2;x}^{(K)}\right)_{11}&=-2K_1\partial_z^2+2\alpha+4\tilde c_{\Omega,\parallel}^{(K)}(z),\nn\\
\left(\hat\Pi_{1,2;x}^{(K)}\right)_{12}&=\left(\hat\Pi_{1,2;x}^{(K)}\right)_{21}=-2\tilde c_{M\Omega}^{(K)}(z),\nn\\
\left(\hat\Pi_{1,2;x}^{(K)}\right)_{22}&=-2K_{123}\partial_z^2+2\alpha+4\tilde c_{M,m_J=\pm 1}^{(K)}(z),
\end{align}
and
\begin{align}
\left(\hat\Pi_{1,2;y}^{(K)}\right)_{11}&=\left(\hat\Pi_{1,2;x}^{(K)}\right)_{11},\nn\\
\left(\hat\Pi_{1,2;y}^{(K)}\right)_{12}&=\left(\hat\Pi_{1,2;y}^{(K)}\right)_{21}=-\left(\hat\Pi_{1,2;x}^{(K)}\right)_{12},\nn\\
\left(\hat\Pi_{1,2;y}^{(K)}\right)_{22}&=\left(\hat\Pi_{1,2;x}^{(K)}\right)_{22}.
\end{align}
One straightforwardly verifies that a uniform $S_x$ or $S_y$ spin wave (i.e., $\varphi=\text{const.}$) costs zero energy, i.e.,
\begin{align}
\hat\Pi_{1,2;x}^{(+)}\left(\begin{array}{c}
\tilde{\Omega}_{1,S_x}^{(+)} \\
\tilde{M}_{3,S_x}^{(+)}
\end{array}\right)=0,\hspace{5mm}
\hat\Pi_{1,2;y}^{(+)}\left(\begin{array}{c}
\tilde{\Omega}_{2,S_y}^{(+)} \\
\tilde{M}_{4,S_y}^{(+)}
\end{array}\right)=0,
\end{align}
if $\Delta_\parallel(z)$ and $\Delta_\perp(z)$ satisfy the GL equations.

Finally, we turn to the Goldstone mode associated with the breaking $SO(2)_{S_z}\times SO(2)_{L_z}\rightarrow SO(2)_{L_z+S_z}$ of relative spin-orbit symmetry. This mode corresponds to an $L_z$ rotation by $\half\varphi(\b{r})$ and a simultaneous $S_z$ rotation by $-\half\varphi(\b{r})$:
\begin{align}
A_{\mu j}(\b{r})=R_{\mu\nu}^{(S)}(\hat{\b{z}},-\half\varphi(\b{r}))R_{jk}^{(L)}(\hat{\b{z}},\half\varphi(\b{r}))\overline{A}_{\nu k}(z).
\end{align}
Expanding for small $\varphi$, we find that this mode is real with $\phi^{(+)}_\text{SO}=\frac{i\varphi}{2}\{L_z,\overline{A}\}=i\varphi\Delta_\parallel L_z$. Thus only the $J=1$, $m_J=0$ component $\Omega_3$ is nonzero, and
\begin{align}
\tilde{\Omega}_{3,\text{SO}}^{(+)}=\varphi\Delta_\parallel.
\end{align}
The $J=1$, $m_J=0$ fluctuation free energy is
\begin{align}
f_{1;0}=\sum_{K=\pm}\int_0^D dz\,\tilde{\Omega}_3^{(K)}\hat{\Pi}_{1;0}^{(K)}\tilde{\Omega}_3^{(K)},
\end{align}
where
\begin{align}
\hat{\Pi}_{1;0}^{(K)}=-2K_1\partial_z^2+2\alpha+4\tilde c_{\Omega,\perp}^{(K)}(z).
\end{align}
One again straightforwardly checks that $\hat{\Pi}_{1;0}^{(+)}\tilde{\Omega}_{3,\text{SO}}^{(+)}=0$ for a uniform fluctuation $\varphi=\text{const.}$, provided that $\Delta_\parallel(z)$ and $\Delta_\perp(z)$ satisfy the GL equations, which identifies $\tilde{\Omega}_{3,\text{SO}}^{(+)}$ as the spin-orbit Goldstone mode.

The remaining contribution to the fluctuation free energy is that of the $J=2$, $m_J=\pm 2$ modes $\tilde{M}_1$ and $\tilde{M}_5$:
\begin{align}
f_{2;\pm 2}=\sum_{K=\pm}\sum_{a=1,5}\int_0^D dz\,\tilde{M}_a^{(K)}\hat\Pi_{2;\pm 2}^{(K)}\tilde{M}_a^{(K)},
\end{align}
where
\begin{align}
\hat\Pi_{2;\pm 2}^{(K)}=-2K_1\partial_z^2+2\alpha+4\tilde c_{M,m_J=\pm 2}^{(K)}(z).
\end{align}
Since we have exhausted the number (4) of Goldstone modes, such modes are necessarily massive.

\subsection{Quadratic term: \texorpdfstring{$\boldsymbol{q}\neq 0$}{Gneq0}}\label{sec:qneq0}

Mizushima and Sauls~\cite{mizushima2018} find that the mode that goes soft at the superfluid crystallization transition is an $m=0^+$ mode, which in our notation belongs to the $J=0,2$, $m_J=0$ multiplet $\{\tilde{\Delta}^{(+)},\tilde{M}_2^{(+)}\}$. We thus focus on the $K=+$ modes in what follows. Having developed the formalism for the ``band-edge'' modes with $\b{q}=0$, we return to the full free energy (\ref{FreeEnergy}), which contains a sum over all in-plane wavevectors $\b{q}$. The real and imaginary modes do not mix at quadratic order, even with $\b{q}\neq 0$. Ignoring the equilibrium piece $f[\overline{A}]$ from now on, Eq.~(\ref{FreeEnergy}) becomes:
\begin{align}\label{FreeEnergyQuad}
f=\sum_\b{q}\int_0^D dz\,\Phi_{-\b{q}}^T(z)\left(\hat{C}(\b{0},z)+\delta\hat{C}(\b{q},z)\right)\Phi_\b{q}(z)+\c{O}(\phi^3).
\end{align}
We define a 9-component vector of fluctuations,
\begin{align}\label{9vec}
\Phi_\b{q}(z)=\left(
\begin{array}{ccccccccc}
\tilde{\Delta}^{(+)} & \tilde{M}_2^{(+)} & \tilde{\Omega}_1^{(+)} & \tilde{M}_3^{(+)} & \tilde{\Omega}_2^{(+)}
& \tilde{M}_4^{(+)} & \tilde{\Omega}_3^{(+)} & \tilde{M}_1^{(+)} & \tilde{M}_5^{(+)}
\end{array}\right)_{\b{G},z}^T.
\end{align}
The $9\times 9$ matrix $\hat{C}(\b{0},z)$ contains all the $\hat{\Pi}^{(+)}_{J;m_J}$ operators of Sec.~\ref{sec:q0}:
\begin{align}\label{C0z}
\hat{C}(\b{0},z)=\left(\begin{array}{cccccc}
\hat{\Pi}_{0,2;0}^{(+)} & & & & & \\
& \hat{\Pi}_{1,2;x}^{(+)} & & & & \\
& & \hat{\Pi}_{1,2;y}^{(+)} & & & \\
& & & \hat{\Pi}_{1;0}^{(+)} & & \\
& & & & \hat{\Pi}_{2;\pm 2}^{(+)} & \\
& & & & & \hat{\Pi}_{2;\pm 2}^{(+)}
\end{array}\right).
\end{align}
Finally, $\delta\hat{C}(\b{q},z)$ contains the $\b{q}$-dependent pieces of the quadratic fluctuation kernel. It is quadratic in $\b{q}$, since the original GL functional (\ref{GL}) is quadratic in spatial derivatives. One can check that it is also explicitly invariant under $SO(2)_{J_z}$ rotations, which act simultaneously on the fields (\ref{9vec}) and on the wavevector $\b{q}$. Explicitly, the nonzero upper-triangular elements are given by:
\begin{align}\label{dCqz}
    \delta\hat{C}_{11}(\b{q},z)&=\delta\hat{C}_{44}(\b{q},z)=\delta\hat{C}_{66}(\b{q},z)=2K_1\b{q}^2,\nn\\
    \delta\hat{C}_{22}(\b{q},z)&=\delta\hat{C}_{77}(\b{q},z)=\delta\hat{C}_{88}(\b{q},z)=\delta\hat{C}_{99}(\b{q},z)=(2K_1+K_{23})\b{q}^2,\nn\\
    \delta\hat{C}_{33}(\b{q},z)&=2(K_1\b{q}^2+K_{23}q_y^2),\nn\\
    \delta\hat{C}_{55}(\b{q},z)&=2(K_1\b{q}^2+K_{23}q_x^2),\nn\\
    \delta\hat{C}_{13}(\b{q},z)&=-\sqrt{2}\delta\hat{C}_{24}(\b{q},z)=\sqrt{2}\delta\hat{C}_{48}(\b{q},z)
    =-\sqrt{2}\delta\hat{C}_{67}(\b{q},z)=-\sqrt{2}\delta\hat{C}_{69}(\b{q},z)=2iK_{23}q_y\partial_z,\nn\\
    -\delta\hat{C}_{15}(\b{q},z)&=-\sqrt{2}\delta\hat{C}_{26}(\b{q},z)=\sqrt{2}\delta\hat{C}_{47}(\b{q},z)
    =-\sqrt{2}\delta\hat{C}_{49}(\b{q},z)=-\sqrt{2}\delta\hat{C}_{68}(\b{q},z)=2iK_{23}q_x\partial_z,\nn\\
    \delta\hat{C}_{28}(\b{q},z)&=-\delta\hat{C}_{79}(\b{q},z)=K_{23}(q_x^2-q_y^2),\nn\\
    \delta\hat{C}_{29}(\b{q},z)&=-\delta\hat{C}_{35}(\b{q},z)=\delta\hat{C}_{78}(\b{q},z)=2K_{23}q_xq_y,
\end{align}
and the lower-triangular elements are given by $\delta\hat{C}_{ji}(\b{q},z)=\delta\hat{C}_{ij}(\b{q},z)$.

\subsection{``\texorpdfstring{$\b{k}\cdot\b{p}$}{kp}'' approach to computing the normal modes}\label{sec:kdotp}

We first review the basic idea of the $\b{k}\cdot\b{p}$ method in electronic bandstructure calculations~\cite{Winkler}, then explain how it can be applied to the diagonalization of the quadratic functional (\ref{FreeEnergyQuad}). This method constructs an effective Bloch Hamiltonian $H_{nn'}(\b{k})$ in the following way. The original Hamiltonian is
\begin{align}
\hat{H}=-\frac{\nabla^2}{2m}+V(\b{r}),
\end{align}
where $V(\b{r})$ is a periodic potential. Using Bloch's theorem, a given eigenstate is of the form $\psi_{\b{k}}(\b{r})=e^{i\b{k}\cdot\b{r}}u_{\b{k}}(\b{r})$ with eigenenergy $E(\b{k})$, where the periodic part of the wavefunction, $u_{\b{k}}(\b{r})$, obeys a $\b{k}$-dependent, reduced Schr\"odinger equation:
\begin{align}\label{redSchrod}
\left(-\frac{\nabla^2}{2m}+V(\b{r})+\frac{k^2}{2m}-\frac{i}{m}\b{k}\cdot\nabla\right)u_{\b{k}}(\b{r})=E(\b{k})u_{\b{k}}(\b{r}).
\end{align}
If we are interested in finding the bandstructure and wavefunctions in the vicinity of $\b{k}=0$, the idea of the $\b{k}\cdot\b{p}$ method is to first solve the $\b{k}=0$ problem,
\begin{align}\label{BandEdgeProblem}
\left(-\frac{\nabla^2}{2m}+V(\b{r})\right)u_{n\b{0}}(\b{r})=E_n(\b{0})u_{n\b{0}}(\b{r}),
\end{align}
which gives us a complete orthonormal set of states $|n\b{0}\rangle$, $u_{n\b{0}}(\b{r})=\langle\b{r}|n\b{0}\rangle$ (the band-edge wavefunctions), with discrete label $n$, since the equation is solved inside a single unit cell with periodic boundary conditions. We use this complete set as a basis to formally expand the unknown solution $u_\b{k}$ at $\b{k}\neq 0$:
\begin{align}\label{FormalExpanduk}
u_\b{k}(\b{r})=\sum_{n'} c_{n'\b{k}}u_{n'\b{0}}(\b{r}).
\end{align}
Substituting this expansion into (\ref{redSchrod}), multiplying on both sides by $u_{n\b{0}}^*(\b{r})$, and integrating over all $\b{r}$, we find that (\ref{redSchrod}) is converted into a matrix eigenvalue problem,
\begin{align}\label{MatEig}
\sum_{n'} H_{nn'}(\b{k})c_{n'\b{k}}=E(\b{k})c_{n\b{k}},
\end{align}
where the effective Bloch Hamiltonian is:
\begin{align}\label{Blocheff}
H_{nn'}(\b{k})=\left(E_n(\b{0})+\frac{k^2}{2m}\right)\delta_{nn'}+\frac{\b{k}}{m}\cdot\langle n\b{0}|-i\nabla|n'\b{0}\rangle,
\end{align}
where we have also used (\ref{BandEdgeProblem}). Knowing the band-edge wavefunctions $u_{n\b{0}}(\b{r})$, we can compute the matrix elements $\langle n\b{0}|-i\nabla|n'\b{0}\rangle=\int d\b{r}\,u_{n\b{0}}^*(\b{r})(-i\nabla)u_{n'\b{0}}(\b{r})$ explicitly, then diagonalize the matrix (\ref{Blocheff}) to obtain the eigenvalues $E(\b{k})$ and eigenvectors $c_{n\b{k}}$, from which we can reconstruct the original Bloch wavefunction via (\ref{FormalExpanduk}). Since the basis $\{u_{n\b{0}}(\b{r})\}$ is really infinite, in practice one keeps only the states with the $N$ lowest band-edge eigenvalues $E_n(\b{0})$, which, at least for sufficiently small $\b{k}$, should give a good approximation to the bandstructure in that range of energies.

We follow essentially the same approach for our problem. The analog of $u_\b{k}(\b{r})$ is $\Phi_\b{q}(z)$, and the analog of the reduced Schr\"odinger equation (\ref{redSchrod}) is:
\begin{align}\label{redSchrodGL}
\left(\hat{C}(\b{0},z)+\delta\hat{C}(\b{q},z)\right)\Phi_\b{q}(z)=\lambda(\b{q})\Phi_\b{q}(z).
\end{align}
The $\b{q}=0$ ``band-edge'' problem is:
\begin{align}
\hat{C}(\b{0},z)\Phi_{n\b{0}}(z)=\lambda_n(\b{0})\Phi_{n\b{0}}(z),
\end{align}
which, given the block-diagonal structure of the $\b{q}=0$ kernel (\ref{C0z}), breaks up into 6 independent systems of (coupled or individual) ordinary differential equations to be solved inside the ``unit cell'' $0<z<D$ with the appropriate Dirichlet/Neumann boundary conditions. In practice, we solve these differential equations by working with a truncated basis of sine ($\propto\sin(n\pi z/D)$) and cosine ($\propto\cos(n\pi z/D)$) functions and diagonalizing the resulting finite-dimensional matrix with standard linear algebra techniques. Finally, in the expansion analogous to (\ref{FormalExpanduk}),
\begin{align}
\Phi_\b{q}(z)=\sum_{n'}c_{n'\b{q}}\Phi_{n'\b{0}}(z),
\end{align}
we also keep a finite number of states $n'$. Proceeding as in the $\b{k}\cdot\b{p}$ problem, we obtain the matrix equation analogous to (\ref{MatEig}):
\begin{align}
\sum_{n'}H_{nn'}(\b{q})c_{n'\b{q}}=\lambda(\b{q})c_{n\b{q}},
\end{align}
where
\begin{align}\label{HG}
H_{nn'}(\b{q})=\lambda_n(\b{0})\delta_{nn'}+\langle n\b{0}|\delta\hat{C}(\b{q},z)|n'\b{0}\rangle,
\end{align}
and we denote
\begin{align}
\langle n\b{0}|\delta\hat{C}(\b{q},z)|n'\b{0}\rangle\equiv\int_0^Ddz\,\Phi_{n\b{0}}^\dag(z)\delta\hat{C}(\b{q},z)\Phi_{n'\b{0}}(z).
\end{align}
Using (\ref{dCqz}) and the sine and cosine basis functions mentioned earlier, the integration over $z$ can be performed analytically, and the matrix elements expressed in terms of the expansion coefficients (obtained numerically) for the ``band-edge wavefunctions'' $\Phi_{n\b{0}}(z)$. Diagonalizing the ``$\b{k}\cdot\b{p}$'' Hamiltonian $H_{nn'}(\b{q})$, we obtain the expansion coefficients $c_{n\b{q}}^{(j)}$ for the eigensolutions $\Phi_\b{q}^{(j)}(z)$ of the Schr\"odinger-like equation (\ref{redSchrodGL}),
\begin{align}
\Phi_\b{q}^{(j)}(z)=\sum_n c_{n\b{q}}^{(j)}\Phi_{n\b{0}}(z),
\end{align}
with eigenvalues $\lambda^{(j)}(\b{q})$. A generic fluctuation $\Phi_\b{q}(z)$ appearing in the free energy (\ref{FreeEnergyQuad}) can be expanded on this basis of eigensolutions:
\begin{align}\label{PhiGexp}
\Phi_\b{q}(z)=\sum_j u^{(j)}_{\b{q}}\Phi_\b{q}^{(j)}(z).
\end{align}
Using $\Phi_{-\b{q}}(z)=\Phi_\b{q}^*(z)$ and the orthonormality of the eigensolutions for each $\b{q}$,
\begin{align}
\int_0^Ddz\,\Phi_\b{q}^{(j)}(z)^\dag\Phi_\b{q}^{(j')}(z)=\delta_{jj'},
\end{align}
the fluctuation free energy density is given by
\begin{align}
f=\sum_\b{q}\sum_j\lambda^{(j)}(\b{q})|u^{(j)}_{\b{q}}|^2+\mathcal{O}(\phi^3),
\end{align}
which is Eq.~(6) in the main text.

\section{Cubic and quartic terms in the PDW functional}

\subsection{Cubic term}\label{sec:cubic}

As discussed earlier, at the quadratic level there is no mixing between real and imaginary modes. The mode $j=j_*$ that goes soft at the transition is a real mode; thus only real modes are kept in the Landau theory for the crystallization transition. At the cubic level, there can be mixing between real and imaginary modes; however, near the crystallization transition ($r\approx 0$ and $|\b{q}|\approx Q$) the imaginary modes have a high free-energy cost and their equilibrium amplitudes are small compared to $u_\b{q}=u^{(j_*)}_{\b{q}}$. Therefore we can justify ignoring the weak cubic mixing between real and imaginary modes. Expanding the original GL functional (\ref{GL}) to cubic order in $\phi$, we have:
\begin{align}\label{f3}
f^{(3)}=8\sum_{abc}\int_0^Ddz\,V_{abc}(\Delta_\parallel,\Delta_\perp)\Phi_a\Phi_b\Phi_c,
\end{align}
with $\Phi$ the 9-component vector (\ref{9vec}), $a,b,c=1,\ldots,9$, and the $V_{abc}$ coefficients are linear functions of the equilibrium order parameters $\Delta_\parallel$ and $\Delta_\perp$. Passing to the Fourier domain, we obtain:
\begin{align}
f^{(3)}=8\sum_{\b{q}_1,\b{q}_2,\b{q}_3}\sum_{abc}\int_0^D dz\,V_{abc}(\Delta_\parallel,\Delta_\perp)\Phi_{\b{q}_1,a}(z)\Phi_{\b{q}_2,b}(z)\Phi_{\b{q}_3,c}(z)\delta_{\b{q}_1+\b{q}_2+\b{q}_3,0}.
\end{align}
Substituting in $\Phi_\b{q}(z)\approx u_\b{q}\Phi_\b{q}^{(j_*)}(z)$, i.e., keeping only the soft mode, we have
\begin{align}
f^{(3)}\approx-\sum_{\b{q}_1,\b{q}_2,\b{q}_3}w(\b{q}_1,\b{q}_2,\b{q}_3)u_{\b{q}_1}u_{\b{q}_2}u_{\b{q}_3}\delta_{\b{q}_1+\b{q}_2+\b{q}_3,0},
\end{align}
where we define
\begin{align}
w(\b{q}_1,\b{q}_2,\b{q}_3)\equiv-8\sum_{abc}\int_0^D dz\,V_{abc}(\Delta_\parallel,\Delta_\perp)\Phi_{\b{q}_1,a}^{(j_*)}(z)\Phi_{\b{q}_2,b}^{(j_*)}(z)\Phi_{\b{q}_3,c}^{(j_*)}(z).
\end{align}
In general, the cubic vertex $w(\b{q}_1,\b{q}_2,\b{q}_3)$ has a nontrivial momentum dependence, which in real space corresponds to a nonlocal interaction $f^{(3)}=-\int d^2r_\parallel\,d^2r_\parallel'\,d^2r_\parallel''\,w(\b{r}_\parallel,\b{r}_\parallel',\b{r}_\parallel'')u(\b{r}_\parallel)u(\b{r}_\parallel')u(\b{r}_\parallel'')$. Following Ref.~\cite{ChaikinLubensky}, we want a simpler theory with a local interaction, $f^{(3)}\approx-w\int d^2r_\parallel\,u^3(\b{r}_\parallel)$. This corresponds to neglecting the momentum dependence of the cubic vertex $w(\b{q}_1,\b{q}_2,\b{q}_3)$. Since the ``wavefunction'' $\Phi_{\b{q}}^{(j_*)}$ is a smooth function of $\b{q}$, we approximate the cubic vertex by its value at zero momentum $\b{q}_1=\b{q}_2=\b{q}_3=0$, which can be viewed as the zeroth-order term in a gradient expansion:
\begin{align}
f^{(3)}\approx-w\sum_{\b{q}_1,\b{q}_2,\b{q}_3}u_{\b{q}_1}u_{\b{q}_2}u_{\b{q}_3}\delta_{\b{q}_1+\b{q}_2+\b{q}_3,0},
\end{align}
which is Eq.~(8) in the main text, where
\begin{align}\label{CubicVertexApprox}
w\equiv w(0,0,0)=-8\sum_{abc}\int_0^D dz\,V_{abc}(\Delta_\parallel,\Delta_\perp)\Phi_{\b{0},a}^{(j_*)}(z)\Phi_{\b{0},b}^{(j_*)}(z)\Phi_{\b{0},c}^{(j_*)}(z).
\end{align}
Since for $\b{q}=0$, $\Phi_\b{0}^{(j_*)}(z)$ is a ``band-edge wavefunction'' with well-defined rotational quantum numbers that belongs to the $\{\tilde{\Delta}^{(+)},\tilde{M}_2^{(+)}\}$ multiplet, only the terms in Eq.~(\ref{CubicVertexApprox}) with $a,b,c$ either 1 or 2 contribute. Explicitly, the relevant $V_{abc}$ coefficients are given by:
\begin{align}
V_{111}=\sqrt{2}\beta_{12345}\Delta_\perp,\hspace{5mm}
V_{112}=2\beta_{12}\Delta_\parallel,\hspace{5mm}
V_{122}=\sqrt{2}\beta_{12}\Delta_\perp,\hspace{5mm}
V_{222}=(2\beta_{12}+\beta_{345})\Delta_\parallel.
\end{align}

We can now understand why $w=0$ at $T_2^*$ but not at $T_1^*$. $T_2^*$ occurs in the planar phase, where $\Delta_\perp=0$ and thus $V_{111}=V_{122}=0$. The critical mode fluctuation in this case is a pure fluctuation of $\Delta_\perp$ about zero, i.e., $\tilde{\Delta}^{(+)}\neq 0$ but $\tilde{M}_2^{(+)}=0$. In other words, only the first component of $\Phi_{\b{0}}^{(j_*)}$ is nonzero, thus (\ref{CubicVertexApprox}) vanishes identically. In the PDB phase with $\Delta_\perp\neq 0$ and $\Delta_\parallel\neq 0$, the critical mode fluctuation has both $\tilde{\Delta}^{(+)}$ and $\tilde{M}_2^{(+)}$ components and $w$ is therefore generically nonzero. However, we must keep in mind that the planar phase is not realized in real $^3$He, thus the $T_2^*$ instability is not physical once strong-coupling corrections are taken into account. The study of the fluctuation spectrum in the A phase with strong-coupling corrections is left for future work.

\subsection{Quartic term}

A quartic term in the PDW functional is also important to ensure thermodynamic stability. This quartic term must have a positive coefficient, otherwise the free energy is unbounded from below and we have to go to higher orders in the Landau expansion. Going back to the original expansion (\ref{OPexpansion}), we see that since the original GL functional (\ref{GL}) is quartic in $A$, the leading contribution to the term quartic in fluctuations $\phi$ is simply obtained by evaluating the quartic term of (\ref{GL}) with $A=\phi$. In the Fourier domain, we have
\begin{align}
f^{(4)}=\sum_{\b{q}_1,\ldots,\b{q}_4}\sum_{abcd}V_{abcd}\int_0^Ddz\,\Phi_{\b{q}_1,a}(z)\Phi_{\b{q}_2,b}(z)
\Phi_{\b{q}_3,c}(z)\Phi_{\b{q}_4,d}(z)\delta_{\b{q}_1+\ldots+\b{q}_4,0},
\end{align}
where $V_{abcd}$ is now constant and only depends on the $\beta_i$ parameters. As for the cubic term, this would in general produce a momentum-dependent vertex $\eta(\b{q}_1,\ldots,\b{q}_4)$. Demanding a local quartic interaction $f^{(4)}\approx\eta\int d^2r_\parallel\,u^4(\b{r}_\parallel)$, we approximate the quartic vertex by its value at $\b{q}_i=0$, $i=1,\ldots,4$:
\begin{align}\label{QuarticVertex}
\eta=\sum_{abcd}V_{abcd}\int_0^Ddz\,\Phi_{\b{0},a}^{(j_*)}(z)\Phi_{\b{0},b}^{(j_*)}(z)\Phi_{\b{0},c}^{(j_*)}(z)\Phi_{\b{0},d}^{(j_*)}(z).
\end{align}
As for the cubic term, $\Phi_{\b{0}}^{(j_*)}(z)$ only has $\tilde{\Delta}^{(+)}$ and $\tilde{M}_2^{(+)}$ components, thus it is sufficient to substitute
\begin{align}
\phi=\Delta^{(+)}\lambda^0+M_2^{(+)}\lambda^2=\frac{1}{\sqrt{3}}\left(\tilde{\Delta}^{(+)}+\sqrt{2}\tilde{M}_2^{(+)}\right)\lambda^0+\frac{1}{\sqrt{3}}\left(\tilde{M}_2^{(+)}-\sqrt{2}\tilde{\Delta}^{(+)}\right)\lambda^2
=\left(\begin{array}{ccc}
\tilde{M}_2^{(+)} & & \\
& \tilde{M}_2^{(+)} & \\
& & \sqrt{2}\tilde{\Delta}^{(+)}
\end{array}\right)
\end{align}
in the quartic term of (\ref{GL}). Note that $\phi=\phi^*$ since only real modes participate (in the approximation where we neglect quartic mixing with high-energy imaginary modes, see Sec.~\ref{sec:cubic}), and also $\phi=\phi^T$, which together imply $\phi=\phi^\dag$. Substituting in (\ref{GL}), we obtain:
\begin{align}
F^{(4)}&=\int d^3r\left[\beta_{12}\left(\tr\phi^2\right)^2+\beta_{345}\tr\phi^4\right]\nn\\
&=\int d^3r\left[4\beta_{12345}\left(\tilde{\Delta}^{(+)}\right)^4+(4\beta_{12}+2\beta_{345})\left(\tilde{M}_2^{(+)}\right)^4+8\beta_{12}\left(\tilde{\Delta}^{(+)}\right)^2\left(\tilde{M}_2^{(+)}\right)^2\right],
\end{align}
thus the only relevant $V_{abcd}$ coefficients in Eq.~(\ref{QuarticVertex}) are
\begin{align}
V_{1111}=4\beta_{12345},\hspace{5mm}
V_{2222}=4\beta_{12}+2\beta_{345},\hspace{5mm}
V_{1122}=8\beta_{12},
\end{align}
which are all positive. Explicitly, we have:
\begin{align}
f^{(4)}\approx\eta\sum_{\b{q}_1,\ldots,\b{q}_4}u_{\b{q}_1}u_{\b{q}_2}u_{\b{q}_3}u_{\b{q}_4}\delta_{\b{q}_1+\ldots+\b{q}_4,0},
\end{align}
which is Eq.~(9) in the main text, with
\begin{align}
\eta=\int_0^Ddz\left[4\beta_{12345}\left(\Phi_{\b{0},1}^{(j_*)}(z)\right)^4
+(4\beta_{12}+2\beta_{345})\left(\Phi_{\b{0},2}^{(j_*)}(z)\right)^4
+8\beta_{12}\left(\Phi_{\b{0},1}^{(j_*)}(z)\right)^2\left(\Phi_{\b{0},2}^{(j_*)}(z)\right)^2\right]>0.
\end{align}
Thus, the PDW free energy functional is stable at quartic order.

\section{Symmetry analysis of the PDW order parameter}\label{sec:symmetry}

In the PDW phase, we set $u_\b{q}=u$ for $\b{q}\in\{\pm\b{G}_1,\pm\b{G}_2,\pm\b{G}_3\}$ and zero otherwise, where we can take $\b{G}_1=Q\hat{x}$, $\b{G}_2=Q(-\frac{1}{2}\hat{x}+\frac{\sqrt{3}}{2}\hat{y})$, and $\b{G}_3=-\b{G}_1-\b{G}_2=Q(-\frac{1}{2}\hat{x}-\frac{\sqrt{3}}{2}\hat{y})$. The superfluid order parameter thus becomes
\begin{align}
A_{\mu j}(\b{r})\approx\overline{A}_{\mu j}(z)+u\sum_{\b{q}=\pm\b{G}_1,\pm\b{G}_2,\pm\b{G}_3}\left(\Phi^{(j_*)}_\b{q}(z)\right)_{\mu j}e^{i\b{q}\cdot\b{r}_\parallel}.
\end{align}
It becomes more practical at this stage to define a new set of real $3\times 3$ matrices $\gamma^1,\ldots,\gamma^9$ as the linear combinations of Gell-Mann matrices (\ref{GellMann}) appropriate to the 9 components (\ref{9vec}) of the superfluid order parameter:
\begin{align}\label{gamma}
\gamma^1&=\frac{1}{\sqrt{3}}(\lambda^0-\sqrt{2}\lambda^2)=\left(\begin{array}{ccc}
0 & & \\
& 0 & \\
& & \sqrt{2}
\end{array}\right),\hspace{5mm}
\gamma^2=\frac{1}{\sqrt{3}}(\lambda^2+\sqrt{2}\lambda^0)=\left(\begin{array}{ccc}
1 & & \\
& 1 & \\
& & 0
\end{array}\right),\hspace{5mm}
\gamma^3=\frac{1}{\sqrt{2}}(i\lambda^6-\lambda^3)=\left(\begin{array}{ccc}
0 & & \\
& 0 & \\
& -\sqrt{2} & 0
\end{array}\right),\nn\\
\gamma^4&=\frac{1}{\sqrt{2}}(\lambda^3+i\lambda^6)=\left(\begin{array}{ccc}
0 & & \\
& 0 & \sqrt{2} \\
& & 0
\end{array}\right),\hspace{5mm}
\gamma^5=\frac{1}{\sqrt{2}}(i\lambda^7+\lambda^4)=\left(\begin{array}{ccc}
& & 0 \\
& 0 & \\
\sqrt{2} & & 
\end{array}\right),\hspace{5mm}
\gamma^6=\frac{1}{\sqrt{2}}(\lambda^4-i\lambda^7)=\left(\begin{array}{ccc}
& & \sqrt{2} \\
& 0 & \\
0 & & 
\end{array}\right),\nn\\
\gamma^7&=i\lambda^8=\left(\begin{array}{ccc}
0 & 1 & \\
-1 & 0 & \\
& & 0
\end{array}\right),\hspace{5mm}
\gamma^8=\lambda^1=\left(\begin{array}{ccc}
1 & & \\
& -1 & \\
& & 0
\end{array}\right),\hspace{5mm}
\gamma^9=\lambda^5=\left(\begin{array}{ccc}
0 & 1 & \\
1 & 0 & \\
& & 0
\end{array}\right).
\end{align}
Focusing on the pure PDW component $\phi_{\mu j}\equiv A_{\mu j}-\overline{A}_{\mu j}$, we have
\begin{align}\label{PDWOP}
\phi_{\mu j}(\b{r})=u\sum_\alpha\gamma_{\mu j}^\alpha\sum_{i=1}^3\left(\Phi^{(j_*)}_{\b{G}_i}(z)_\alpha e^{i\b{G}_i\cdot\b{r}_\parallel}+\Phi_{-\b{G}_i}^{(j_*)}(z)_\alpha e^{-i\b{G}_i\cdot\b{r}_\parallel}\right),
\end{align}
where $\Phi^{(j_*)}_\b{q}(z)_1,\ldots,\Phi^{(j_*)}_\b{q}(z)_9$ are the 9 components of (\ref{9vec}).

In practice, we calculate $\Phi^{(j_*)}_\b{q}(z)$ numerically for a single wavevector $\b{q}=Q\hat{x}=\b{G}_1$, and obtain the other $\Phi^{(j_*)}_{\b{G}_i}(z)$ and $\Phi_{-\b{G}_i}^{(j_*)}(z)$ by symmetry. First, $\Phi_{\pm\b{G}_2}^{(j_*)}(z)$ and $\Phi_{\pm\b{G}_3}^{(j_*)}(z)$ can be obtained from $\Phi_{\pm\b{G}_1}^{(j_*)}(z)$ by $C_6$ rotations. $\hat{C}(\b{q},z)$ possesses a continuous $SO(2)_{J_z}$ rotation symmetry, expressed as
\begin{align}\label{RdagCR}
\c{R}^{-1}(\theta)\hat{C}(R_\theta\b{q},z)\c{R}(\theta)=\hat{C}(\b{q},z),
\end{align}
where
\begin{align}
R_\theta=\left(\begin{array}{cc}
\cos\theta & -\sin\theta \\
\sin\theta & \cos\theta
\end{array}\right),
\end{align}
is the ordinary $SO(2)$ rotation matrix, and $\c{R}(\theta)=e^{-i\theta\c{J}_z}$ is a 9-dimensional representation of $SO(2)$ with generator
\begin{align}\label{curlyJz}
\c{J}_z=\left(\begin{array}{cc}
0 & 0 \\
0 & 0
\end{array}\right)\oplus
\left(\begin{array}{cccc}
& & -i & \\
& & & i \\
i & & & \\
& -i & &
\end{array}\right)\oplus
\left(\begin{array}{c}
0
\end{array}\right)\oplus
\left(\begin{array}{cc}
& -2i \\
2i &
\end{array}\right).
\end{align}
Explicitly, we have
\begin{align}\label{curlyR}
\c{R}(\theta)=\left(\begin{array}{cc}
1 &  \\
 & 1
\end{array}\right)\oplus
\left(\begin{array}{cccc}
\cos\theta & 0 & -\sin\theta & 0 \\
0 & \cos\theta & 0 & \sin\theta \\
\sin\theta & 0 & \cos\theta & 0 \\
0 & -\sin\theta & 0 & \cos\theta
\end{array}\right)\oplus
\left(\begin{array}{c}
1
\end{array}\right)\oplus
\left(\begin{array}{cc}
\cos 2\theta & -\sin 2\theta \\
\sin 2\theta & \cos 2\theta
\end{array}\right).
\end{align}
Equation (\ref{RdagCR}) implies that if $\Phi_\b{q}^{(j)}(z)$ is an eigenstate of $\hat{C}(\b{q},z)$ with eigenvalue $\lambda^{(j)}(\b{q})$, then $\c{R}(\theta)\Phi_\b{q}^{(j)}(z)$ is an eigenstate of $\hat{C}(R_\theta\b{q},z)$ with the same eigenvalue. Since $\b{G}_2=R_{2\pi/3}\b{G}_1$ and $\b{G}_3=R_{4\pi/3}\b{G}_1$, we can thus take $\Phi^{(j_*)}_{\b{G}_i}(z)=\c{R}(\theta_i)\Phi^{(j_*)}_{\b{G}_1}(z)$, $i=1,2,3$, with $\theta_1=0$, $\theta_2=2\pi/3$, and $\theta_3=4\pi/3$. Likewise, $\Phi^{(j_*)}_{-\b{G}_i}(z)=\c{R}(\theta_i)\Phi^{(j_*)}_{-\b{G}_1}(z)$. Equation (\ref{PDWOP}) thus becomes
\begin{align}\label{PDWOP2}
\phi_{\mu j}(\b{r})=u\sum_\alpha\gamma_{\mu j}^\alpha\sum_{i=1}^3\c{R}_{\alpha\beta}(\theta_i)\left(\Phi_{\b{G}_1}^{(j_*)}(z)_\beta e^{i\b{G}_i\cdot\b{r}_\parallel}+\Phi_{-\b{G}_1}^{(j_*)}(z)_\beta e^{-i\b{G}_i\cdot\b{r}_\parallel}\right).
\end{align}
Next, $\Phi^{(j_*)}_{-\b{G}_1}(z)$ is related to $\Phi^{(j_*)}_{\b{G}_1}(z)$ by a reflection symmetry $x\rightarrow-x$. One can check explicitly that
\begin{align}\label{Mx}
\c{M}_x^{-1}\hat{C}(-q_x,q_y;z)\c{M}_x=\hat{C}(q_x,q_y;z),
\end{align}
where $\c{M}_x=\diag(1,1,1,1,-1,-1,-1,1,-1)$. In particular, $\c{M}_x^{-1}\hat{C}(-\b{G}_1,z)\c{M}_x=\hat{C}(\b{G}_1,z)$, thus we can choose $\Phi_{-\b{G}_1}^{(j_*)}(z)=\c{M}_x\Phi_{\b{G}_1}^{(j_*)}(z)$. Since $\c{M}_x$ is diagonal, Eq.~(\ref{PDWOP2}) becomes
\begin{align}\label{PDWOP3}
\phi_{\mu j}(\b{r})=u\sum_{\alpha\beta}\gamma_{\mu j}^\alpha\sum_{i=1}^3\c{R}_{\alpha\beta}(\theta_i)
\left(e^{i\b{G}_i\cdot\b{r}_\parallel}+\c{M}_x^{\beta\beta}e^{-i\b{G}_i\cdot\b{r}_\parallel}\right)
\Phi_{\b{G}_1}^{(j_*)}(z)_\beta.
\end{align}
Since $\c{M}_x^{\beta\beta}=\pm 1$, the $\b{r}_\parallel$ dependence of the PDW order parameter will be via linear combinations of $\cos\b{G}_i\cdot\b{r}_\parallel$ and $\sin\b{G}_i\cdot\b{r}_\parallel$.

Further simplifications occur due to the fact that $\hat{C}(\b{G}_1,z)$ should be invariant under a $y\rightarrow-y$ symmetry, since $\b{G}_1$ is entirely along the $x$ axis. Analogously to Eq.~(\ref{Mx}), one can check that
\begin{align}
\c{M}_y^{-1}\hat{C}(q_x,-q_y;z)\c{M}_y=\hat{C}(q_x,q_y;z),
\end{align}
where $\c{M}_y=\diag(1,1,-1,-1,1,1,-1,1,-1)$. In particular, $[\c{M}_y,\hat{C}(\b{G}_1,z)]=0$, thus $\Phi^{(j_*)}_{\b{G}_1}(z)$ must be an eigenvector of $\c{M}_y$ with eigenvalue $\pm 1$, since $\c{M}_y^2=1$. Since $\c{M}_y$ is diagonal, we can easily write down its action on the 9 components of $\Phi^{(j_*)}_{\b{G}_1}(z)$. Denoting $\Phi\equiv\Phi^{(j_*)}_{\b{G}_1}(z)$ for simplicity, we have
\begin{align}\label{My}
(\c{M}_y\Phi)_{1,2,5,6,8}=\Phi_{1,2,5,6,8},\hspace{5mm}
(\c{M}_y\Phi)_{3,4,7,9}=-\Phi_{3,4,7,9}.
\end{align}
There are two possibilities. If $\c{M}_y\Phi=\Phi$, then $\Phi_{3,4,7,9}$ must all vanish. Conversely, if $\c{M}_y\Phi=-\Phi$, then $\Phi_{1,2,5,6,8}$ must all vanish. We find numerically that near the PDW instability, the $\c{M}_y$ eigenvalue of $\Phi$ is $+1$, thus $\Phi_{3,4,7,9}=0$.

\subsection{Structure of the PDW order parameter: group-theoretic analysis}

By assuming $u_\b{q}=u$ for all $\b{q}\in\{\pm\b{G}_1,\pm\b{G}_2,\pm\b{G}_3\}$, we have constructed an order parameter $\phi$ that is invariant under the action of a spin-orbital point group $D_6^{L_z+S_z}$ which acts simultaneously on spin and orbital coordinates:
\begin{align}\label{D6Jzaction}
    \c{D}(g)\phi(g^{-1}\b{r}_\parallel,z)\c{D}^{-1}(g)=\phi(\b{r}_\parallel,z),
\end{align}
where $g$ is any element of the point group $D_6$ of the triangular lattice. In the main text, we express $\phi$ in the form
\begin{align}
    \phi=u\sum_{j,k}\phi^{(j,k)}(z)X^{(j,k)}(\b{r}_\parallel),
\end{align}
where $X^{(j,k)}(\b{r}_\parallel)$ is the $k$th $D_6^{L_z+S_z}$ invariant---i.e., a $3\times 3$ matrix transforming according to Eq.~(\ref{D6Jzaction})---associated with the irreducible representation (irrep) $j$ of $D_6$, and $\phi^{(j,k)}(z)$ is the corresponding nonuniversal amplitude (a scalar function of $z$). We now explain how the $X$ invariants can be constructed~\cite{BirPikus} and give a precise definition of the amplitudes plotted in Fig.~4(a).

The dihedral group $D_6$ is a finite group of order 12 consisting of 6 rotations $C_6^0=e,C_6,\ldots,C_6^5$ and 6 reflections $\sigma_1,\sigma_2,\sigma_3,\sigma_1',\sigma_2',\sigma_3'$. We define $C_6$ as a counterclockwise rotation by $\pi/3$, $\sigma_1,\sigma_2,\sigma_3$ as the reflections that leave $\b{G}_1,\b{G}_2,\b{G}_3$ invariant, respectively, and $\sigma_1',\sigma_2',\sigma_3'$ as the reflections that map $\b{G}_1$ to $\b{G}_2$, $\b{G}_2$ to $\b{G}_3$, and $\b{G}_3$ to $-\b{G}_1$, respectively. $D_6$ has 2 generators, which we take to be $C_6$ and $\sigma_1$, and 6 conjugacy classes, which is also the number of irreps. The theory of invariants is based on the fact that the trivial representation only appears in the direct product of two mutually conjugate representations~\cite{BirPikus}; thus invariant functions can be formed only from the product of basis functions belonging to a representation and its conjugate. To enumerate all possible invariants, we need look only at all possible irreps of $D_6$, which are the $A_1$, $A_2$, $B_1$, $B_2$, $E_1$, and $E_2$ irreps. (These irreps are all real and thus self-conjugate.) In the present single-harmonic approximation, the PDW order parameter (\ref{PDWOP}) is expressed as a linear combination of 6 scalar $\b{r}_\parallel$-dependent functions
\begin{align}\label{psi}
    \psi_1=\cos\b{G}_1\cdot\b{r}_\parallel,\hspace{5mm}
    \psi_2=\sin\b{G}_1\cdot\b{r}_\parallel,\hspace{5mm}
    \psi_3=\cos\b{G}_2\cdot\b{r}_\parallel,\hspace{5mm}
    \psi_4=\sin\b{G}_2\cdot\b{r}_\parallel,\hspace{5mm}
    \psi_5=\cos\b{G}_3\cdot\b{r}_\parallel,\hspace{5mm}
    \psi_6=\sin\b{G}_3\cdot\b{r}_\parallel,\hspace{5mm}
\end{align}
multiplying 9 matrices $\gamma^1,\ldots,\gamma^9$, with $z$-dependent coefficients. The 6 functions (\ref{psi}) transform according to a (reducible) 6-dimensional representation of $D_6$, and the 9 matrices (\ref{gamma}) according to a (reducible) 9-dimensional representation. To construct the $X$ invariants, we must first decompose those reducible representations into a direct sum of irreps and construct irreducible basis functions that transform according to the latter. Then, we may form invariant products for each irrep.

We begin with the 6-dimensional representation $M$ according to which the $\psi_j(\b{r}_\parallel)$, $j=1,\ldots,6$ transform:
\begin{align}
    \psi_j(g^{-1}\b{r}_\parallel)=\sum_i\psi_i(\b{r}_\parallel)M_{ij}(g),\hspace{5mm}g\in D_6.
\end{align}
$M$ can be defined by its expression on the generators,
\begin{align}
    M(C_6)=\left(\begin{array}{cccccc}
    0 & 0 & 1 & 0 & 0 & 0 \\
    0 & 0 & 0 & -1 & 0 & 0 \\
    0 & 0 & 0 & 0 & 1 & 0 \\
    0 & 0 & 0 & 0 & 0 & -1 \\
    1 & 0 & 0 & 0 & 0 & 0 \\
    0 & -1 & 0 & 0 & 0 & 0
    \end{array}\right),\hspace{10mm}
    M(\sigma_1)=\left(\begin{array}{cccccc}
    1 & 0 & 0 & 0 & 0 & 0 \\
    0 & 1 & 0 & 0 & 0 & 0 \\
    0 & 0 & 0 & 0 & 1 & 0 \\
    0 & 0 & 0 & 0 & 0 & 1 \\
    0 & 0 & 1 & 0 & 0 & 0 \\
    0 & 0 & 0 & 1 & 0 & 0
    \end{array}\right).
\end{align}
We next need the representation matrices $D^{(j)}(g)$ for the 6 irreps $j=A_1,A_2,B_1,B_2,E_1,E_2$ of $D_6$. These are given by:
\begin{align}\label{Dirreps}
    D^{(A_1)}(C_6)&=1,\hspace{5mm}D^{(A_1)}(\sigma_1)=1;\nn\\
    D^{(A_2)}(C_6)&=1,\hspace{5mm}D^{(A_2)}(\sigma_1)=-1;\nn\\
    D^{(B_1)}(C_6)&=-1,\hspace{5mm}D^{(B_1)}(\sigma_1)=1;\nn\\
    D^{(B_2)}(C_6)&=-1,\hspace{5mm}D^{(B_2)}(\sigma_1)=-1;\nn\\
    D^{(E_1)}(C_6)&=\left(\begin{array}{cc}
    \frac{1}{2} & -\frac{\sqrt{3}}{2} \\
    \frac{\sqrt{3}}{2} & \frac{1}{2}
    \end{array}\right),\hspace{5mm}
    D^{(E_1)}(\sigma_1)=\left(\begin{array}{cc}
    1 & 0 \\
    0 & -1
    \end{array}\right);\nn\\
    D^{(E_2)}(C_6)&=\left(\begin{array}{cc}
    -\frac{1}{2} & -\frac{\sqrt{3}}{2} \\
    \frac{\sqrt{3}}{2} & -\frac{1}{2}
    \end{array}\right),\hspace{5mm}
    D^{(E_2)}(\sigma_1)=\left(\begin{array}{cc}
    1 & 0 \\
    0 & -1
    \end{array}\right).
\end{align}
The reducible representation $M$ decomposes as the direct sum $M=\bigoplus_j a_jD^{(j)}=A_1\oplus B_1\oplus E_1\oplus E_2$, where the $a_j$ are obtained from $a_j=\frac{1}{h}\sum_{g\in G}\chi^{(j)}(g)^*\chi^{(M)}(g)$, where $h=12$ is the order of the group $G=D_6$, $\chi^{(j)}$ is the character of irrep $j$, and $\chi^{(M)}$ is the character of $M$. To construct basis functions belonging to a given irrep $j$, we use the fact that for fixed $\nu=1,\ldots,d_j$ with $d_j=\dim D^{(j)}$, the $6\times 6$ matrices
\begin{align}
    \Pi_{\mu\nu}^{(j)}=\frac{d_j}{h}\sum_{g\in G}D_{\mu\nu}^{(j)}(g)^*M(g),\hspace{5mm}\mu=1,\ldots,d_j,
\end{align}
project a function onto irrep $j$. For simplicity, we set $\nu=1$ and consider the matrices
\begin{align}
    P_\mu^{(j)}=\frac{d_j}{h}\sum_{g\in G}D_{\mu 1}^{(j)}(g)^*M(g),\hspace{5mm}\mu=1,\ldots,d_j.
\end{align}
These matrices are projectors in the sense that for an arbitrary 6-component vector $\b{f}$, the vectors $P_\mu^{(j)}\b{f}$, $\mu=1,\ldots,d_j$ transform among themselves according to irrep $j$:
\begin{align}\label{projtrans}
    M(g)P_\nu^{(j)}\b{f}=\sum_\mu\left(P_\mu^{(j)}\b{f}\right)D_{\mu\nu}^{(j)}(g),\hspace{5mm}g\in D_6.
\end{align}
Using this, we easily construct the following unit vectors,
\begin{align}
    \hat{\b{e}}^{(A_1)}&=\frac{1}{\sqrt{3}}(1,0,1,0,1,0),\nn\\
    \hat{\b{e}}^{(B_1)}&=\frac{1}{\sqrt{3}}(0,1,0,1,0,1),\nn\\
    \hat{\b{e}}_1^{(E_1)}&=\frac{1}{\sqrt{6}}(0,2,0,-1,0,-1),\hspace{5mm}\hat{\b{e}}_2^{(E_1)}=\frac{1}{\sqrt{2}}(0,0,0,1,0,-1),\nn\\
    \hat{\b{e}}_1^{(E_2)}&=\frac{1}{\sqrt{6}}(2,0,-1,0,-1,0),\hspace{5mm}\hat{\b{e}}_2^{(E_2)}=\frac{1}{\sqrt{2}}(0,0,-1,0,1,0),
\end{align}
which transform as $M(g)\hat{\b{e}}_\nu^{(j)}=\sum_\mu\hat{\b{e}}_\mu^{(j)}D_{\mu\nu}^{(j)}(g)$ as per Eq.~(\ref{projtrans}), where the $\mu,\nu$ indices can be suppressed for one-dimensional irreps. Note that the $P_\mu^{(j)}$ vanish identically for $A_2$ and $B_2$ since those irreps do not appear in the direct-sum decomposition of $M$. Using those unit vectors, we can construct irreducible basis functions $\Psi_\mu^{(j)}(\b{r}_\parallel)=\hat{\b{e}}_\mu^{(j)}\cdot\boldsymbol{\psi}(\b{r}_\parallel)$, where $\boldsymbol{\psi}=(\psi_1,\ldots,\psi_6)$ is given in Eq.~(\ref{psi}). These irreducible basis functions transform as
\begin{align}
    \Psi_\nu^{(j)}(g^{-1}\b{r}_\parallel)=\sum_\mu\Psi_\mu^{(j)}(\b{r}_\parallel)D_{\mu\nu}^{(j)}(g),\hspace{5mm}g\in D_6.
\end{align}

We now repeat this exercise for the 9-dimensional representation $\Gamma$ according to which the matrices $\gamma^\alpha$, $\alpha=1,\ldots,9$ transform:
\begin{align}
    \c{D}(g)\gamma^\beta\c{D}^{-1}(g)=\sum_\alpha\gamma^\alpha\Gamma_{\alpha\beta}(g),\hspace{5mm}g\in D_6.
\end{align}
Here $\c{D}$ is itself a 3-dimensional representation of $D_6$ consisting of $O(3)\supset D_6$ matrices; its $SO(3)$ subgroup corresponds to spin-orbital rotations discussed in Eq.~(\ref{SOrotation}), thus $\c{D}(C_6)=e^{-i\pi\lambda^8/3}$. Since $\sigma_1$ is a reflection in the $y$ direction, we thus have:
\begin{align}
    \c{D}(C_6)=\left(\begin{array}{ccc}
    \frac{1}{2} & -\frac{\sqrt{3}}{2} & 0 \\
    \frac{\sqrt{3}}{2} & \frac{1}{2} & 0 \\
    0 & 0 & 1
    \end{array}\right),\hspace{10mm}
    \c{D}(\sigma_1)=\left(\begin{array}{ccc}
    1 & 0 & 0 \\
    0 & -1 & 0 \\
    0 & 0 & 1
    \end{array}\right).
\end{align}
Comparing with Eq.~(\ref{Dirreps}), it is clear that $\c{D}=E_1\oplus A_1$, with the $A_1$ factor corresponding to the trivial action of $D_6$ on the $z$ coordinate. Turning to the $\Gamma$ representation, we have, in the notation of Eqs.~(\ref{curlyJz}-\ref{curlyR}), $\Gamma(C_6)=e^{-i\pi\c{J}_z/3}=\c{R}(\pi/3)$, and by explicit computation we obtain $\Gamma(\sigma_1)=\diag(1,1,-1,-1,1,1,-1,1,-1)=\c{M}_y$. This reducible representation decomposes as $\Gamma=\bigoplus_j a_jD^{(j)}=A_1\oplus A_1\oplus A_2\oplus E_1\oplus E_1\oplus E_2$, where the multiplicities are obtained from $a_j=\frac{1}{h}\sum_{g\in G}\chi^{(j)}(g)^*\chi^{(\Gamma)}(g)$. To construct irreducible basis functions (which in this case will be $3\times 3$ basis matrices), we again form the projectors
\begin{align}
    P_\mu^{(j)}=\frac{d_j}{h}\sum_{g\in G}D_{\mu 1}^{(j)}(g)^*\Gamma(g).
\end{align}
Because of the two-fold multiplicities of the $A_1$ and $E_1$ irreps inside $\Gamma$, we expect there will be two sets of $A_1$ and $E_1$ basis matrices. This expresses itself in the fact that for those irreps, the action of $P_\mu^{(j)}$ on a generic 9-component vector $\b{f}$ will give a linear combination of two linearly independent basis vectors. In this way we construct the following unit vectors:
\begin{align}\label{basisvecs9D}
    \hat{\b{e}}^{(A_1,1)}&=(1,0,0,0,0,0,0,0,0),\nn\\
    \hat{\b{e}}^{(A_1,2)}&=(0,1,0,0,0,0,0,0,0),\nn\\
    \hat{\b{e}}^{(A_2)}&=(0,0,0,0,0,0,1,0,0),\nn\\
    \hat{\b{e}}_1^{(E_1,1)}&=(0,0,0,0,1,0,0,0,0),\hspace{5mm}\hat{\b{e}}_2^{(E_1,1)}=(0,0,-1,0,0,0,0,0,0),\nn\\
    \hat{\b{e}}_1^{(E_1,2)}&=(0,0,0,0,0,1,0,0,0),\hspace{5mm}\hat{\b{e}}_2^{(E_1,2)}=(0,0,0,1,0,0,0,0,0),\nn\\
    \hat{\b{e}}_1^{(E_2)}&=(0,0,0,0,0,0,0,1,0),\hspace{5mm}\hat{\b{e}}_2^{(E_2)}=(0,0,0,0,0,0,0,0,1),
\end{align}
which transform as $\Gamma(g)\hat{\b{e}}_\nu^{(j)}=\sum_\mu\hat{\b{e}}_\mu^{(j)}D_{\mu\nu}^{(j)}(g)$, where the $\mu,\nu$ indices are again suppressed for one-dimensional irreps. Here $\hat{\b{e}}^{(A_1,1)}$ and $\hat{\b{e}}^{(A_1,2)}$ denote two independent basis vectors transforming in the $A_1$ irrep, and likewise $\hat{\b{e}}_\mu^{(E_1,1)}$ and $\hat{\b{e}}_\mu^{(E_1,2)}$ are two independent sets of basis vectors transforming in the $E_1$ irrep. Also note that $P_\mu^{(B_1)}=P_\mu^{(B_2)}=0$ since those irreps do not appear in the direct-sum decomposition of $\Gamma$. From the unit vectors (\ref{basisvecs9D}) and $\boldsymbol{\gamma}\equiv(\gamma^1,\ldots,\gamma^9)$ we form irreducible basis matrices $\Lambda_\mu^{(j)}=\hat{\b{e}}_\mu^{(j)}\cdot\boldsymbol{\gamma}$, which transform as
\begin{align}
    \c{D}(g)\Lambda_\nu^{(j)}\c{D}^{-1}(g)=\sum_\mu\Lambda_\mu^{(j)}D_{\mu\nu}^{(j)}(g),\hspace{5mm}g\in D_6.
\end{align}

From the irreducible basis functions $\Psi_\mu^{(j)}(\b{r}_\parallel)$ and irreducible basis matrices $\Lambda_\mu^{(j)}$, we may finally form $D_6$ invariants via
\begin{align}
    X^{(j,\alpha\beta)}(\b{r}_\parallel)=\sum_{\mu=1}^{d_j}\Lambda_\mu^{(j,\alpha)}\Psi_\mu^{(j,\beta)}(\b{r}_\parallel)^*,
\end{align}
where $\Lambda_\mu^{(j,\alpha)}$ is the $\alpha$th irreducible basis matrix belonging to irrep $j$, and $\Psi_\mu^{(j,\beta)}(\b{r}_\parallel)$ is the $\beta$th irreducible basis function belonging to irrep $j$. (Since the irreps and associated basis functions are real, we may in fact drop the complex conjugate.) We note that only the $A_1$, $E_1$, and $E_2$ irreps appear in both the direct-sum decompositions of $M$ and $\Gamma$; thus only for those irreps can invariants be constructed. Explicitly, the relevant irreducible basis functions are thus:
\begin{align}\label{irrbasisfcts}
    \Psi^{(A_1)}(\b{r}_\parallel)&=\hat{\b{e}}^{(A_1)}\cdot\boldsymbol{\psi}=\frac{1}{\sqrt{3}}\left(\cos Qx+2\cos\textstyle\frac{1}{2}Qx\cos\frac{\sqrt{3}}{2}Qy\right),\nn\\
    \boldsymbol{\Psi}^{(E_1)}(\b{r}_\parallel)&=\left(\hat{\b{e}}_1^{(E_1)}\cdot\boldsymbol{\psi},\hat{\b{e}}_2^{(E_1)}\cdot\boldsymbol{\psi}\right)
    =\sqrt{\frac{2}{3}}\left(\sin Qx+\sin\textstyle\frac{1}{2}Qx\cos\frac{\sqrt{3}}{2}Qy,\sqrt{3}\cos\frac{1}{2}Qx\sin\frac{\sqrt{3}}{2}Qy\right),\nn\\
    \boldsymbol{\Psi}^{(E_2)}(\b{r}_\parallel)&=\left(\hat{\b{e}}_1^{(E_2)}\cdot\boldsymbol{\psi},\hat{\b{e}}_2^{(E_2)}\cdot\boldsymbol{\psi}\right)
    =\sqrt{\frac{2}{3}}\left(\cos Qx-\cos\textstyle\frac{1}{2}Qx\cos\frac{\sqrt{3}}{2}Qy,-\sqrt{3}\sin\frac{1}{2}Qx\sin\frac{\sqrt{3}}{2}Qy\right),
\end{align}
which can be understood as $s$-wave, $p$-wave, and $d$-wave triangular lattice harmonics, respectively (see, e.g., Ref.~\cite{zhou2008}, where slightly different linear combinations are chosen). Likewise, the $A_1,E_1,E_2$ irreducible basis matrices are:
\begin{align}\label{irrbasismat}
    \Lambda^{(A_1,1)}&=\hat{\b{e}}^{(A_1,1)}\cdot\boldsymbol{\gamma}=\gamma^1,\nn\\
    \Lambda^{(A_1,2)}&=\hat{\b{e}}^{(A_1,2)}\cdot\boldsymbol{\gamma}=\gamma^2,\nn\\
    \boldsymbol{\Lambda}^{(E_1,1)}&
    =\left(\hat{\b{e}}_1^{(E_1,1)}\cdot\boldsymbol{\gamma},\hat{\b{e}}_2^{(E_1,1)}\cdot\boldsymbol{\gamma}\right)
    =(\gamma^5,-\gamma^3),\nn\\
    \boldsymbol{\Lambda}^{(E_1,2)}&
    =\left(\hat{\b{e}}_1^{(E_1,2)}\cdot\boldsymbol{\gamma},\hat{\b{e}}_2^{(E_1,2)}\cdot\boldsymbol{\gamma}\right)
    =(\gamma^6,\gamma^4),\nn\\
    \boldsymbol{\Lambda}^{(E_2)}&
    =\left(\hat{\b{e}}_1^{(E_2)}\cdot\boldsymbol{\gamma},\hat{\b{e}}_2^{(E_2)}\cdot\boldsymbol{\gamma}\right)
    =(\gamma^8,\gamma^9).
\end{align}
Therefore, only 5 invariants can be constructed: $\Lambda^{(A_1,1)}\Psi^{(A_1)}$, $\Lambda^{(A_1,2)}\Psi^{(A_1)}$, $\b{\Lambda}^{(E_1,1)}\cdot\b{\Psi}^{(E_1)}$, $\b{\Lambda}^{(E_1,2)}\cdot\b{\Psi}^{(E_1)}$, and $\b{\Lambda}^{(E_2)}\cdot\b{\Psi}^{(E_2)}$. In fact, using the explicit form of the matrices $\c{R}(\theta_i)$ and $\c{M}_x$, we can show that the PDW order parameter (\ref{PDWOP3}) is a linear combination of those 5 invariants:
\begin{align}\label{PDWOP4}
\phi_{\mu j}(\b{r})&=2\sqrt{3}u\left(\Phi_1(z)\Lambda^{(A_1,1)}_{\mu j}+\Phi_2(z)\Lambda^{(A_1,2)}_{\mu j}\right)\Psi^{(A_1)}(\b{r}_\parallel)\nn\\
&\phantom{=}+i\sqrt{6}u\left(\Phi_5(z)\b{\Lambda}^{(E_1,1)}_{\mu j}+\Phi_6(z)\b{\Lambda}^{(E_1,2)}_{\mu j}\right)\cdot\b{\Psi}^{(E_1)}(\b{r}_\parallel)\nn\\
&\phantom{=}+\sqrt{6}u\Phi_8(z)\b{\Lambda}^{(E_2)}_{\mu j}\cdot\b{\Psi}^{(E_2)}(\b{r}_\parallel),
\end{align}
where we recall the notation $\Phi_\beta(z)\equiv\Phi_{\b{G}_1}^{(j_*)}(z)_\beta$ employed earlier, see Eq.~(\ref{My}).

\begin{figure}[t]
    \includegraphics[width=0.5\columnwidth]{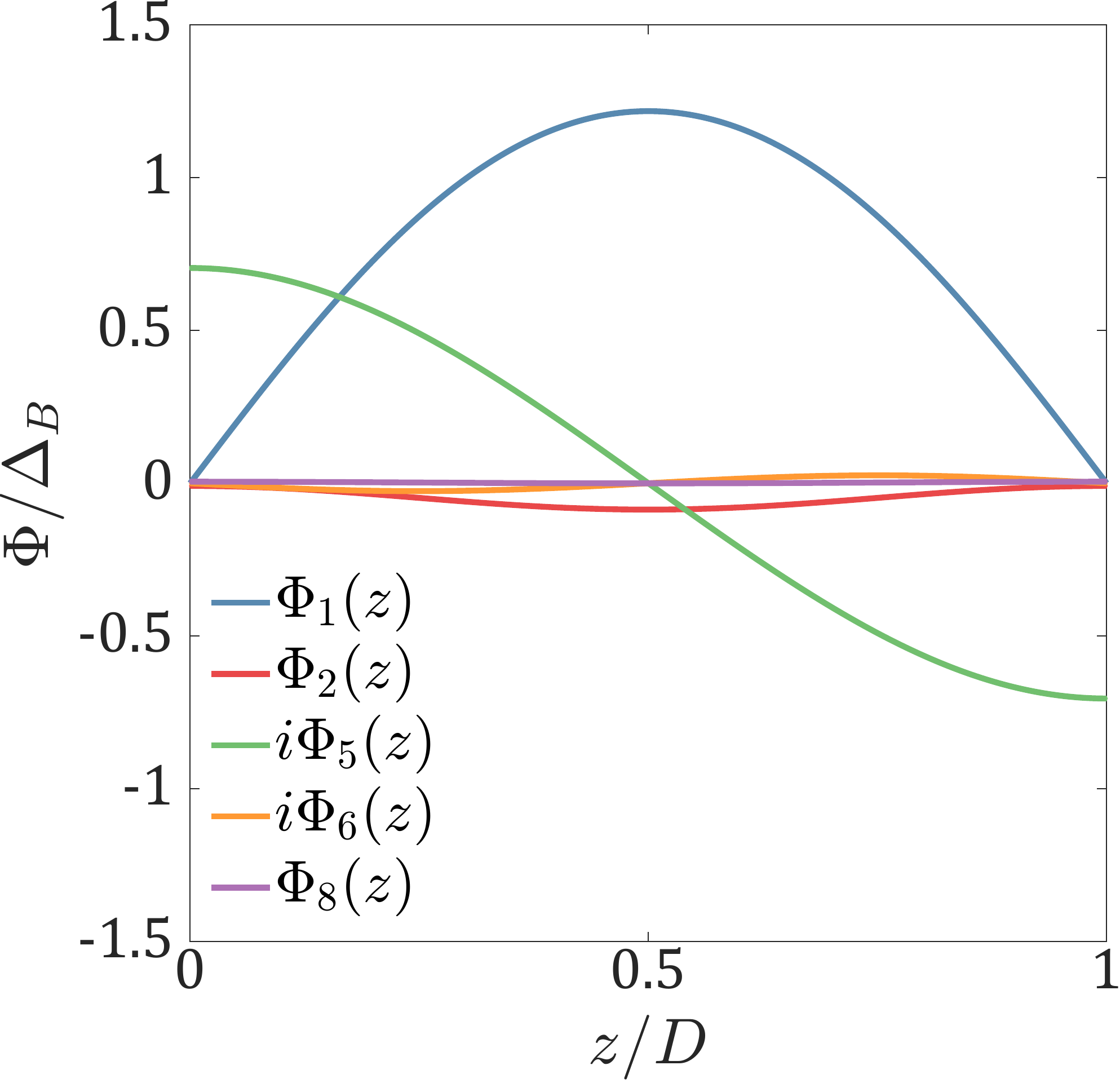}
    \caption{Irreducible $z$-dependent amplitudes of the PDW order parameter.}
    \label{fig:zamps}
    \centering
\end{figure}

Finally, we discuss the consequences of time-reversal symmetry. The ``Hamiltonian'' $\hat{C}(\b{q},z)$ has the property $\hat{C}(\b{q},z)^*=\hat{C}(-\b{q},z)$ as can be checked explicitly. This implies the spectrum of $\hat{C}$ is symmetric under $\b{q}\rightarrow-\b{q}$. Since $\Phi_{\b{G}_1}^{(j_*)}(z)$ is nondegenerate in the region of interest (near and at the instability), we can thus choose $\Phi_{-\b{G}_1}^{(j_*)}(z)=\Phi_{\b{G}_1}^{(j_*)}(z)^*$. However, we saw earlier that reflection symmetry in the $x$ direction implies that we can choose $\Phi_{-\b{G}_1}^{(j_*)}(z)=\c{M}_x\Phi_{\b{G}_1}^{(j_*)}(z)$. Combining these two equalities, and using the explicit diagonal form of $\c{M}_x$ presented earlier, we find that $\Phi_{1,2,3,4,8}$ can be made real (in a suitable choice of gauge), while $\Phi_{5,6,7,9}$ is purely imaginary. Together with the earlier numerical observation that $\Phi_{3,4,7,9}=0$, we find that $\Phi_{1,2,8}$ is real and $\Phi_{5,6}$ purely imaginary, which can be confirmed numerically. Since the basis functions (\ref{irrbasisfcts}) and basis matrices (\ref{irrbasismat}) are real, this implies that the PDW order parameter (\ref{PDWOP4}) is real for real $u$, as expected. The real amplitudes $\Phi_1(z)$, $\Phi_2(z)$, $i\Phi_5(z)$, $i\Phi_6(z)$, and $\Phi_8(z)$ are plotted in Fig.~4(a) of the main text and also reproduced here in Fig.~\ref{fig:zamps}. To aid in the visualization of the three-dimensional structure of the PDW, we also give animated plots of the nine components $\phi_{11}(\b{r})$, $\phi_{12}(\b{r})$, $\ldots$, $\phi_{33}(\b{r})$ of the PDW order parameter (\ref{PDWOP4}) in units of $u$ in a set of nine ancillary files \texttt{phi11.gif}, \texttt{phi12.gif}, $\ldots$, \texttt{phi33.gif}.

\bibliography{3He}